\documentclass[aps,prd,amssymb,nofootinbib,twocolumn,epsf,floatfix]{revtex4}
\usepackage[usenames]{color} 
\usepackage{graphicx}
\usepackage{amssymb}
\usepackage{amsmath}
\usepackage{epstopdf}
\usepackage{latexsym}
\usepackage{bm}

\newcommand{\half}{{\scriptstyle\frac{1}{2}}}

\newcommand{\DeltaI}{\Delta^{\kern -0.2em(1)}}
\newcommand{\DeltaII}{\Delta^{\kern -0.2em(2)}}
\newcommand{\DeltaN}{\Delta^{\kern -0.2em(n)}}
\newcommand{\deltaI}{\delta^{(1)}}
\newcommand{\deltaII}{\delta^{(2)}}
\newcommand{\deltaN}{\delta^{(n)}}
\newcommand{\deltaNR}{\delta_N}
\newcommand{\deltaINR}{\delta^{(1)}_N}
\newcommand{\deltaIINR}{\delta^{(2)}_N}

\newcommand{\deltaRR}{\delta_R}
\newcommand{\deltaIRR}{\delta^{(1)}_R}
\newcommand{\deltaIIRR}{\delta^{(2)}_R}

\newcommand{\tdeltaIRR}{\widetilde\delta^{(1)}_R}

\newcommand{\be}{\begin{equation}}  
\newcommand{\ee}{\end{equation}}  
\newcommand{\bea}{\begin{eqnarray}}  
\newcommand{\eea}{\end{eqnarray}}  
 
\newcommand{\na}{\nabla} 
\newcommand{\Lie}{\mbox{\pounds}} 
\def\up#1{\raise1mm\hbox{$\!\!^{#1}$}} 
\def\upp#1{\raise2mm\hbox{$\!\!^{#1}$}} 
\def\Upp#1{\raise2mm\hbox{$\!\!\!\!^{#1}\,\,$}} 
\def\uppp#1{\raise2.5mm\hbox{$\!\!^{#1}$}}

\newcommand{\bsube}{\begin{subequations}}
\newcommand{\esube}{\end{subequations}}

\begin{document}
 
\title{Differential rotation of the unstable nonlinear r-modes}

\author{John L. Friedman${}^1$,  Lee Lindblom${}^{2,3,4}$, and  
Keith H. Lockitch${}^5$}

\affiliation{${}^1$Leonard Parker Center for Gravitation, Cosmology and Astrophysics,
Department of Physics, University of Wisconsin-Milwaukee, P.O. Box 413, 
Milwaukee, Wisconsin 53201, USA} 
\email{friedman@uwm.edu}

\affiliation{${}^2$Theoretical Astrophysics 350-17, California Institute of
Technology, Pasadena, CA 91125, USA}

\affiliation{${}^3$Center for Astrophysics and Space Sciences 0424, University 
of California at San Diego, 9500 Gilman Drive, La Jolla, CA 92093-0424, USA}

\affiliation{${}^4$Mathematical Sciences Center, Tsinghua University,
Beijing 100084, China}

\affiliation{${}^5$Center for Theoretical Astrophysics, Department
of Physics, University of Illinois at Urbana-Champaign, 
Urbana IL 61801, USA}
\altaffiliation{Now at The Ayn Rand Institute, 2121 Alton Parkway, Irvine, CA 92606, USA}

\date{\today}
 
\begin{abstract}
At second order in perturbation theory, the $r$-modes of uniformly rotating 
stars include an axisymmetric part that can be identified with  
differential rotation of the background star. If one does not include 
radiation-reaction, the differential rotation is constant in time and 
has been computed by S\'a. It has a gauge dependence associated 
with the family of time-independent perturbations that add differential 
rotation to the unperturbed equilibrium star: For stars with a barotropic 
equation of state, one can add to the time-independent second-order solution 
arbitrary differential rotation that is stratified on cylinders (that is a 
function of distance $\varpi$ to the axis of rotation). We show here that 
the gravitational radiation-reaction force that drives the $r$-mode 
instability removes this gauge freedom:  The exponentially
growing differential 
rotation of the unstable second-order $r$-mode is unique.  We derive a general
expression for this rotation law for Newtonian models and evaluate it
explicitly for slowly rotating models with polytropic equations of
state.
\end{abstract}
 
\maketitle

\section{Introduction}

Unstable $r$-modes \cite{A97,FM98} may limit the angular velocity of 
old neutron stars spun up by accretion and may contribute to the spin-down of 
nascent neutron stars (see \cite{Lindblom98b,fsbook,Bondarescu07,Bondarescu09} 
for references and reviews).    Spruit \cite{Spruit99} argued that 
angular momentum loss from the star would generate differential rotation, 
because the loss rate depends on the mode shape and varies over the star. 
Growing differential rotation winds up and amplifies the star's magnetic field, 
and Rezzolla and collaborators~\cite{Rezzolla00,Rezzolla01b,Rezzolla01c}, studied 
the possibility that the energy lost to the magnetic field would damp out 
the $r$-mode instability. (In Spruit's scenario, a buoyancy instability of 
the greatly enhanced magnetic field could power a $\gamma$-ray burst.) 
To estimate the magnetic-field windup, Rezzolla {\it et al.} used a 
drift velocity of a fluid element; this is second-order in perturbation 
theory, but because the second-order velocity field had not been computed, 
they estimated it by integrating the first order velocity field.  Subsequently,  
Cuofano {\it et al.} \cite{Cuofano2010,Cuofano_etal12} used this estimate of drift 
velocity to study the evolution of the $r$-mode instability damped by magnetic 
field wind-up.%
\footnote{Work by Abbassi, {\it et al.}~\cite{ARR12} 
also looks at the damping of 
$r$-modes due to a magnetic field; here, however, the magnetic dissipation 
arises from magnetic diffussivity in a linearized MHD treatment.}

Following Spruit's work, Levin and Ushomirsky found the differential rotation 
of the unstable $r$-mode in a toy model of a spherical shell of fluid 
\cite{LevinUshormirsky01}.  S\'a \cite{Sa2004} then carried out the first 
computation of the differential rotation associated with a stable $r$-mode of 
uniformly rotating barotropic Newtonian stellar models and, with collaborators, 
looked at implications of the calculation for the unstable mode~\cite{Sa2005a,Sa2005b}.  
The differential rotation arises at second order in perturbation theory as a 
time-independent, axisymmetric part of the solution to the perturbed Euler 
equations; for the $r$-mode whose linear part is associated with the 
angular harmonic $Y^{\ell\ell}$, S\'a's solution has the form
\be
   \delta^{(2)}\Omega 
  =  \alpha^2 \Omega C_\Omega 
\left(\frac zR\right)^2\left(\frac \varpi R\right)^{2\ell-4}
	+ \alpha^2 \delta^{(2)}_N\Omega(\varpi).
\label{e:sa_rotation}\ee
Here $\alpha$ measures the amplitude of the first-order perturbation,
$C_\Omega$ is dimensionless and of order unity, the $z$-axis is
the axis of rotation, and $\varpi$ is the distance from the axis.  The
function $\delta^{(2)}_N\Omega(\varpi)$ is arbitrary.  This ambiguity in the
rotation law is present for the following reason. One can perturb a
uniformly rotating barotropic star by adding differential rotation,
changing the angular velocity from $\Omega$ to
$\Omega+\delta\Omega(\varpi)$.  If $\delta\Omega(\varpi)$ is chosen to
be quadratic in $\alpha$, $\delta\Omega(\varpi)=\alpha^2
\delta^{(2)}_N\Omega(\varpi)$, it and the corresponding time-independent
perturbations of density, pressure, and gravitational potential
$\Phi$, constitute a solution to the time-independent second-order
perturbation equations.  Cao {\it et al.}~\cite{CZW15} use a particular 
choice of $\delta^{(2)}\Omega$ to recompute the magnetic damping.  

In the present paper, we show that the second-order radiation-reaction
force removes the ambiguity in the differential rotation associated
with the Newtonian $r$-modes.  In effect, the degeneracy in the space of
zero-frequency solutions is broken by the radiation-reaction force,
which picks out a unique differential rotation law that depends on the
neutron-star equation of state.  We find an explicit formula for that
rotation law for the unstable $r$-modes of
slowly rotating stars.  

To lowest nonvanishing post-Newtonian order, the growth time $\tau$ of the 
radiation-reaction driven (CFS) instability of an $r$-mode is given by  
\[
 \beta\equiv \frac1\tau 
	= C_\beta \frac G{c^{2\ell+3}} M R^{2\ell} \Omega^{2\ell+2}, 
\] 
where $C_\beta$ is a dimensionless constant that depends on the
equation of state.  In using the Newtonian Euler equation together
with the radiation-reaction force at lowest nonvanishing
post-Newtonian order, we are neglecting radiation-reaction terms
smaller by factors of ${\cal O}(R\Omega/c)$ and ${\cal O}(GM/Rc^2)$; this means, in
particular, that we keep only terms linear in the dimensionless
parameter $\beta/\Omega$.

Three small parameters appear in the paper: The amplitude $\alpha$ of 
the perturbation, the dimensionless growth rate $\beta/\Omega$, and, 
in the final, slow-rotation part of the paper, the angular velocity $\Omega$. 
For the logic of the paper, it is helpful to 
note that these three parameters can be regarded as independent of one another.  
The growth rate $\beta$ can be varied by changing the equation of state of
the material while keeping $\alpha$ and $\Omega$ fixed; for example, in polytropes
(stars based on the polytropic equation of state $p=K\rho^n$), one can change
$\beta$ by changing the polytropic constant $K$.   

The plan of the paper is as follows.  Sect.~\ref{s:newtonian} lists
the equations governing a Newtonian star acted on by a post-Newtonian
radiation-reaction force, with the star modeled as a self-gravitating
perfect fluid.  In Sect.~\ref{s:Perturbed Stellar Models}, we discuss
first- and second-order perturbations of a uniformly rotating star.
From the second-order equations, we obtain a formal expression for the
unique differential rotation law of an unstable $r$-mode in
terms of the first-order perturbations and second-order contributions 
that will turn out to be of higher-order in $\Omega$. 
Up to this point in the paper, the analysis holds
for rapidly rotating stars.  In Sect.~\ref{s:SlowRotation}, we
specialize to a slowly rotating background, keeping terms of lowest
nonvanishing order in $\Omega$ and thereby obtaining an explicit 
formula for the radiation-reaction induced differential rotation.
Finally, a discussion section briefly comments on the validity of 
the results for an accreting neutron star, when one includes magnetic fields, 
nonzero initial data for other modes, and viscosity.   

Our notation for fluid perturbations is chosen to make explicit the 
orders of the expansions in the amplitude $\alpha$ and angular 
velocity $\Omega$. The notation is
defined as it is introduced in Secs. II and III, but, for easy
reference, we also provide a table that summarizes the notation
in Appendix~\ref{s:Notation}.  We use gravitational units, setting $G=c=1$.

\section{Newtonian Stellar Models}
\label{s:newtonian}

Let $Q=\{\rho, v^a,p,\Phi\}$ 
denote the collection of fields that
determine the state of the fluid in a self-gravitating Newtonian
stellar model.  The quantity $\rho$ represents the mass density, $v^a$
the fluid velocity, $p$ the pressure, and $\Phi$ the gravitational
potential.  For a barotropic equation of state $p=p(\rho)$, 
the specific enthalpy $h$ of the fluid is  
\be
h= \int_0^p \frac{dp}{\rho},
\ee
and we define a potential $U$ by
\be
U= h + \Phi.
\ee
The evolution of the fluid is determined by Euler's equation,
the mass-conservation law, and the Poisson equation for the Newtonian gravitational
potential.  These equations may be written as
\bea
E^a&\equiv& \partial_tv^a + v^b\nabla_bv^a + \nabla^a U=f^a_{GR},
\label{e:FullEulerEquation}\\
0&=&\partial_t\rho + \nabla_a(\rho v^a),\\
\nabla^2\Phi &=& 4\pi\rho.
\label{e:GravPotentialEq}
\eea

The version of the Euler equation that we use, 
Eq.~(\ref{e:FullEulerEquation}), includes $\vec f_{GR}$, the
post-Newtonian gravitational radiation-reaction force (per unit mass).
This force plays a central role in the nonlinear evolution of the
$r$-modes that is the primary focus of our paper. It is 
given by
\bea
&&
\!\!\!\!\!
\vec f_{GR}=\sum_{l\geq 2}\sum_{|m|\leq l} \frac{(-1)^{\ell+1}N_\ell}{32\pi}
\,\Re\Biggl\{
\frac{\vec \nabla(r^\ell Y^{\ell m})}{\sqrt{\ell}}
\frac{d^{\,2\ell+1}I^{\ell m}}{dt^{\,2\ell+1}}
\nonumber\\
&&\!\!\!\!\!
-
\frac{2r^\ell\vec Y^{\ell m}_{B}}{\sqrt{\ell+1}}
\frac{d^{\,2\ell+2}S^{\ell m}}{dt^{\,2\ell+2}}
-
\frac{2\vec v\times \vec \nabla(r^\ell Y^{\ell m})}{\sqrt{\ell}}
\frac{d^{\,2\ell+1}S^{\ell m}}{dt^{\,2\ell+1}}\Biggr\},\quad
\label{e:GRForceDef}
\eea
where $\Re (Z)$ denotes the real part of a complex quantity $Z$.  The
quantities $I^{\ell m}$ and $S^{\ell m}$ are the complex mass and
current multiple moments of the fluid source
(cf. Thorne~\cite{Thorne1980} Eqs.~5.18a,b) defined by,
\bea
I^{\ell m}&=&
\frac{N_\ell}{\sqrt{\ell}} 
\int \rho\, r^\ell Y^{*\ell m} d^3 x,\\
S^{\ell m}&=&\frac{2N_\ell}{\sqrt{\ell+1}}
\int \rho \,r^\ell \vec v \cdot \vec Y^{*\ell m}_{B} d^3x,
\eea
with $N_\ell$ the constant
\bea
N_\ell = \frac{16\pi}{(2\ell+1)!!}
\sqrt{\frac{(\ell+2)(\ell+1)}{2(\ell-1)}}.
\label{e:NlDef}
\eea
The functions $Y^{\ell m}$ are the standard spherical harmonics,
while the $\vec Y^{\ell m}_B$ are the magnetic-type vector
harmonics defined by
\bea
\vec Y^{\ell m}_B = \frac{\vec r\times \vec\nabla Y^{lm}}{\sqrt{\ell(\ell+1)}}.
\eea
We use the normalizations $1=\int |Y^{\ell m}|^2 d\cos\theta d\phi$
and $1=\int |\vec Y^{\ell m}_B|^2 d\cos\theta d\phi$ for these
spherical harmonics.  In Cartesian coordinates $\vec r$ is given by
$\vec r =(x,y,z)$. We point out that this expression for the gravitational
radiation-reaction force, Eq.~(\ref{e:GRForceDef}), agrees with the
mass-multipole part of the force given by Ipser and
Lindblom~\cite{Ipser1991}.  It also agrees with the current-multipole
part of the force given by Lindblom, {\it et al.}~\cite{Lindblom2001}
(following Blanchet~\cite{Blanchet1997} and Rezzolla, {\it et
al.}~\cite{Rezzolla1999}) for the $\ell=2$ and $m=2$ case.  The general
form of the force given in Eq.~(\ref{e:GRForceDef}), however, is new.
 
The post-Newtonian radiation-reaction force is gauge dependent, so the
expression for it is not unique.  We derived the expression for
  the force given in Eq.~(\ref{e:GRForceDef}) by
  requiring that it implies a time-averaged (over several
  oscillation periods) power $\langle\!\langle
  dE/dt\rangle\!\rangle|_{GR}$ (which is gauge invariant), and angular
  momentum flux $\langle\!\langle d\vec J/dt\rangle\!\rangle|_{GR}$
  lost to gravitational waves that agree with the standard
  post-Newtonian expressions, cf. Thorne~\cite{Thorne1980}.  We
  present expressions for these flux quantities in
  Appendix~\ref{s:RadiationReaction} that are equivalent to, but are
  somewhat simpler than the standard ones.

We consider small perturbations of rigidly rotating, axisymmetric,
barotropic equilibrium models (models with a barotropic equation 
of state).  The fluid velocity in these equilibria is 
denoted
\be 
\vec v = \Omega\,\vec\phi,
\label{e:EquilibriumV}
\ee 
where $\vec\phi$ generates rotations about the $z$ axis; in Cartesian
coordinates, $\vec\phi =(-y,x,0)$.  For barotropic equilibria, 
Euler's equation reduces to 
\be
0= \nabla_a(h+\Phi -\half \varpi^2\Omega^2),
\label{e:EquilibriumEuler}
\ee
where $h$ is the specific enthalpy of the fluid and $\varpi$ is the cylindrical
radial coordinate, $\varpi^2=x^2+y^2$.  
The surface of the star is the boundary where the
pressure and the enthalpy vanish: $p=h=0$.

\section{Perturbed Stellar Models}
\label{s:Perturbed Stellar Models}

We denote by $Q(\alpha,t,\vec x)$ a one-parameter family of stellar
models. For each value of the parameter $\alpha$, $Q(\alpha,t,\vec x)$ 
satisfies the full nonlinear time-dependent 
Eqs.~(\ref{e:FullEulerEquation})--(\ref{e:GravPotentialEq}).  
We assume that the model with $\alpha=0$ is an axisymmetric 
equilibrium model, as described in 
Eqs.~(\ref{e:EquilibriumV})--(\ref{e:EquilibriumEuler}).
The exact perturbation $\delta Q$, defined as the difference between
$Q(\alpha)$ and $Q(0)$, is defined everywhere on the intersection of
the domains where $Q(\alpha)$ and $Q(0)$ are defined:
\be
\delta Q(\alpha, t,\vec x)\equiv Q(\alpha,t,\vec x)-Q(0,t,\vec x).
\ee
It is also be useful to define $\deltaN Q$, the derivatives
of the one parameter family $Q(\alpha)$ evaluated at the
unperturbed stellar model, where $\alpha=0$:
\be
\deltaN Q(t,\vec x) = \frac{1}{n!}\frac{\partial^{\,n}\, 
Q(\alpha,t,\vec x)}
{\partial\alpha^n}\biggr|_{\alpha=0}.
\label{e:deltaN Q Def}
\ee
These derivatives can be used to define a formal power series expansion
for $\delta Q$:
\be
\delta Q(\alpha, t,\vec x)= \alpha\, \deltaI Q(t,\vec x) + \alpha^2\, \deltaII
Q(t,\vec x) + {\cal O}(\alpha^3).
\label{e:delta Q}
\ee
Each point in 
the interior of the unperturbed star is, for sufficiently small $\alpha$, 
in the interior of the perturbed star; the derivatives 
$\delta^{(n)} Q$ defined in Eq.~(\ref{e:deltaN Q Def}) 
and the formal power series expansion in Eq.~(\ref{e:delta Q}) 
are thus well-defined at all points of the interior of 
the unperturbed star, but may diverge at the surface.
We consider constant-mass sequences of stellar models, i.e., models
whose exact mass perturbations, $\delta M = M(\alpha)- M(\alpha=0)$
vanish identically for all values of
$\alpha$.  The integrals of the $n^\mathrm{th}$-order density
perturbations therefore vanish identically for these models:
\be
0=\frac{1}{n!}\left.\frac{d^{\,n} M(\alpha)}{d\alpha^n}\right|_{\alpha=0} =
\int \deltaN \rho\, \sqrt{g}\,d^{\,3}x.
\label{e:MassConservationIntegral}
\ee

  The exact (to all orders in the perturbation parameter
$\alpha$) perturbed evolution equations for these stellar
models can be written in the form
\bea
\delta E^a &=& (\partial_t  + \Omega \Lie_\phi) \delta v^a
+ 2 \Omega \delta v^b \nabla_b\phi^a
+\nabla^a\delta U,\nonumber\\
&&\qquad\qquad\qquad\,\,
+ \,\,\delta v^b\nabla_b\delta v^a = \delta f_{GR}^a,
\label{e:PerturbedEulerExact}\\
0&=&(\partial_t  + \Omega \Lie_\phi)\delta\rho 
+\nabla_a(\rho\,\delta v^a + \delta \rho\,\delta v^a),
\label{e:PerturbedMassConsExact}\qquad\\
\nabla^2\delta\Phi &=&4\pi\delta\rho,
\label{e:PerturbedGravPotentialEqExact}
\eea
where $\Lie_\phi$ is the Lie derivative along the vector field $\vec \phi$, 
and $\rho$ is the density of the unperturbed star.
The exact perturbed gravitational radiation-reaction force $\delta \vec f_{GR}$
that appears in Eq.~(\ref{e:PerturbedEulerExact}) is given by
\bea
&&
\!\!\!
\delta \vec f_{GR}=\sum_{l\geq 2}\sum_{|m|\leq l} \frac{(-1)^{\ell+1}N_\ell}{32\pi}
\,\Re\Biggl\{
\frac{\vec \nabla(r^\ell Y^{\ell m})}{\sqrt{\ell}}
\frac{d^{\,2\ell+1}\delta I^{\ell m}}{dt^{\,2\ell+1}}
\nonumber\\
&&
-
\frac{2r^\ell\vec Y^{\ell m}_{B}}{\sqrt{\ell+1}}
\frac{d^{\,2\ell+2}\delta S^{\ell m}}{dt^{\,2\ell+2}}
-
\frac{2\Omega\vec\phi
\times \vec \nabla(r^\ell Y^{\ell m})}{\sqrt{\ell}}
\frac{d^{\,2\ell+1}\delta S^{\ell m}}{dt^{\,2\ell+1}}\nonumber\\
&&
-
\frac{2\delta\vec v
\times \vec \nabla(r^\ell Y^{\ell m})}{\sqrt{\ell}}
\frac{d^{\,2\ell+1}\delta S^{\ell m}}{dt^{\,2\ell+1}}
\Biggr\},
\label{e:PerturbedGRForceExact}
\eea
where 
%
\bea
\delta I^{\ell m}&=&
\frac{N_\ell}{\sqrt{\ell}} 
\int \delta \rho\, r^\ell Y^{*\ell m} d^3 x,
\label{e:PerturbedMassMultipole}\\
\delta S^{\ell m}&=&\frac{2N_\ell}{\sqrt{\ell+1}}
\int r^\ell \left[\rho\,\delta \vec v+\delta \rho \,
\left(\Omega\vec\phi
+\delta \vec v\right)\right]\cdot \vec Y^{*\ell m}_{B} d^3x,
\nonumber
\label{e:PerturbedCurrentMultipole}\\
\eea

   It is convenient to decompose the perturbations 
$\delta Q$ into parts $\deltaNR Q$ that satisfy the pure Newtonian
evolution equations, and parts $\deltaRR Q$ caused by the addition of
the gravitational radiation-reaction force.  In particular the
nonradiative stellar perturbations $\deltaNR Q$ satisfy the perturbed
Euler equation:
\be
\delta \vec E = 0.
\ee
When the effects of gravitational radiation-reaction are included, the
complete perturbation, $\delta Q$, satisfies the Euler equation
driven by the gravitational radiation-reaction force
\be
\delta \vec E = \delta \vec f_{GR}.
\ee

\subsection{First Order Perturbations}
\label{s:FirstOrderPerturbations}

The classical first-order (in powers of $\alpha$) $r$-modes have
angular and temporal dependence \cite{PP78,fsbook}
\bea \deltaINR \rho &=& \deltaINR \hat\rho_- \,\sin \psi_N,
\label{e:deltaINRrho}\\ 
\deltaINR v^a &=& \varpi^{-2}\phi^a\phi_b\deltaINR \hat v^b_+\, 
\sin \psi_N +P^{\,a}{}_b
\deltaINR \hat v^b_+\, \cos \psi_N,\nonumber\\\\ 
\deltaINR U &=& \deltaINR \hat U_-\,\sin \psi_N,\\ 
\deltaINR \Phi &=& \deltaINR \hat \Phi_-\,\sin \psi_N,
\label{e:deltaINRPhi} 
\eea 
where $\psi_N=\omega_Nt+m\phi$, with $m\neq 0$.  The tensor
\be
 P^{\,a}{}_b\equiv \delta^a{}_b-\varpi^{-2}\phi^a\phi_b 
\label{e:projection}\ee
is the
projection operator orthogonal to $\phi^a$, and $\deltaINR \hat
Q=\deltaINR \hat Q(\varpi,z)$ depends on the cylindrical coordinates
$\varpi$ and $z$, but not on $\phi$ or $t$.  The origin of time has
been chosen to give the perturbations definite parity under the
diffeomorphism $\phi\rightarrow -\phi$ at $t=0$.  We use the term
$\phi$-{\it parity} to mean parity under this transformation.  The
subscripts $\pm$ indicate that $\deltaINR\hat\rho_-$, $\deltaINR
\hat U_-$, and $\deltaINR \hat\Phi_-$ are parts of odd $\phi$-parity
scalars, while $\deltaINR \hat v^a_+$ is part of an even $\phi$-parity
vector field.

When gravitational radiation reaction is included, the Euler equation
is altered by the relatively weak radiation-reaction force $\vec f_{GR}$.
The first order radiation-reaction force can be written in the form:
\bea
\deltaI \vec f_{GR}=\beta\deltaINR \vec v_+ + \delta^{(1)}_\perp \vec f_{GR+}, 
\label{e:GRForceParityEq}
\eea
where $\beta$ is the growth rate of the $r$-mode instability, and
$\delta^{(1)}_\perp \vec f_{GR+}$ is (by definition)
the even $\phi$-parity part of the
radiation-reaction force that is orthogonal to $\deltaINR \vec v_+$ 
and that 
therefore does not contribute directly to the energy evolution of the mode.  
Equation~(\ref{e:PerturbedGRForceExact}) implies that the odd
$\phi$-parity part of the radiation-reaction force, $\delta^{(1)}_\perp \vec f_{GR-}$, 
vanishes when the classical $r$-mode is chosen to have the
$\phi$-parity given in Eqs.~(\ref{e:deltaINRrho})--(\ref{e:deltaINRPhi}).
The gravitational radiation-reaction force causes an instability by
introducing an imaginary part $\beta$ to the frequency of the
mode. The overall structure of the modes is therefore changed in the
following way (schematically):
\bea \deltaI \rho &=& \left(\deltaINR \hat\rho_-+\deltaIRR \hat\rho_-\right) 
\sin\psi \,e^{\beta t}
+\deltaIRR \hat\rho_+
\cos\psi\, e^{\beta t},
\label{e:deltaIrho}\nonumber\\\\ 
\deltaI v^a &=& \deltaIRR \hat v^b_-
\Bigl[\varpi^{-2}\phi^a\phi_b\, \cos\psi+P^{\,a}{}_b
\, \sin\psi\Bigr]e^{\beta t}\nonumber\\ 
&&+
\left(\deltaINR \hat v^b_++\deltaIRR \hat v^b_+\right)
\times\nonumber\\
&&\quad\Bigl[\varpi^{-2}\phi^a\phi_b\, \sin\psi+P^{\,a}{}_b
\, \cos\psi\Bigr]e^{\beta t},\\
\deltaI U &=&
\left(\deltaINR \hat U_-+\deltaIRR \hat U_-\right)
\sin\psi\, e^{\beta t}\nonumber\\
&&\quad+\deltaIRR \hat U_+
\cos\psi\, e^{\beta t},\\ 
\deltaI \Phi &=&\left(
\deltaINR \hat \Phi_-+\deltaIRR \hat \Phi_-\right)
\sin\psi\, e^{\beta t}\nonumber\\
&&\quad+\deltaIRR \hat \Phi_+
\cos\psi\, e^{\beta t}, 
\label{e:deltaIPhi}
\eea where $\psi=\psi_N+\psi_R=\omega_N t +\omega_R t + m\phi$.  The
radiative corrections $\deltaIRR \hat Q$ are smaller than the
nonradiative perturbations $\deltaINR\hat Q$ by terms of order ${\cal
  O}(\beta/\omega_N)$.  The radiative correction $\omega_R$ to the frequency,
is smaller than $\omega_N$ by a term of order ${\cal  O}(\beta/\omega_N)^2$, 
so we ignore that change here, setting $\psi=\psi_N$.%
\footnote{Friedman and Schutz~\cite{FriedmanSchutz1975}
  derive the following general expression for the frequencies of the
  modes of Lagrangian systems (including Newtonian fluids with
  gravitational radiation-reaction forces):
  $0=A(\omega+i\beta)^2-(B+iD)(\omega+i\beta)-C$, where $A$, $B$, $C$
  and $D$ are real.  The term $D$ vanishes for non-dissipative
  Newtonian fluid stars.  When $D$ is small, it is straightforword to
  show that the real part of the frequency, $\omega$, differs from the
  frequency of the non-dissipative $D=0$ system, $\omega_N$, by terms
  of order $D^2$: $\omega=\omega_N+{\cal O}(D^2)$.  It is also easy to
  show that the imaginary part of the frequency $\beta$ is
  proportional to $D$ for a mode with $\beta_N=0$.}

The radiative corrections to the $r$-mode, $\deltaIRR Q$, are
determined by substituting
Eqs.~(\ref{e:deltaIrho})--(\ref{e:deltaIPhi}) into the first-order
perturbed mass conservation and Euler equations.  After applying the
equations satisfied by the nonradiative parts of the perturbations,
$\deltaINR Q$, the resulting system of equations can be divided into
parts proportional to $\sin\psi_N$ and $\cos\psi_N$ respectively, each of
which must vanish separately.  The resulting equations can be divided
further into a set that determines $\deltaIRR \hat\rho_- $, $\deltaIRR
\hat U_- $, and $\deltaIRR\hat v^a_+$, and another that determines
$\deltaIRR \hat \rho_+ $, $\deltaIRR \hat U_+ $, and $\deltaIRR\hat
v^a_-$.

The equations that determine the radiative corrections
having the same $\phi$-parity as the classical nonradiative $r$-modes
are then
\bea 
&& \!\!\!\!\!
(\omega_N+m\Omega)\,\deltaIRR \hat\rho_- 
+ m\rho\varpi^{-2} \phi_a\,\deltaIRR\hat v^a_+\nonumber\\
&&\qquad
+\nabla_a\left(\rho P^a{}_b\deltaIRR \hat v^b_+\right)=0,\quad 
\label{e:FirstOrderOddParityRhoEq}\\
&&\!\!\!\!\!
\left[(\omega_N+m\Omega)\phi_a+2\varpi\Omega\nabla_a\varpi\right]
\deltaIRR\hat v^a_+= - m\,\deltaIRR \hat U_-,\quad\\ 
&&\!\!\!\!\!
\left[(\omega_N+m\Omega)P^a{}_b+\frac2\varpi\Omega\nabla^a\varpi\phi_b\right]
\deltaIRR\hat v^b_+\nonumber\\
&&\qquad=  P^{ab}\nabla_b\,\deltaIRR \hat U_-. 
\label{e:FirstOrderEvenParityVelocityEq}
\eea
These equations are homogeneous and are identical to those satisfied
by the classical $r$-modes.  The solutions for $\deltaIRR \hat \rho_-
$, $\deltaIRR \hat U_- $, and $\deltaIRR\hat v^a_+$ are therefore
proportional to the classical $r$-modes: $\deltaINR \hat \rho_- $,
$\deltaINR \hat U_- $, and $\deltaINR\hat v^a_+$.  The effect of
adding these radiative corrections to the classical $r$-modes is
simply to re-scale its amplitude.  We choose to keep the amplitude,
$\alpha$, of the mode fixed, and therefore without loss of generality
we set
\bea
0=\deltaIRR\hat \rho_- =\deltaIRR \hat U_- =\deltaIRR\hat v^a_+.
\eea

It follows that the first-order radiative corrections have $\phi$-parity 
opposite to that of the classical $r$-modes: $\deltaIRR
\hat\rho=\deltaIRR\hat\rho_+$, $\deltaIRR \hat U = \deltaIRR \hat U_+
$, and $\deltaIRR\hat v^a = \deltaIRR\hat v^a_-$.  They are determined
by the equations
\bea
&&\!\!\!\!\!
 (\omega_N+m\Omega)\,\deltaIRR\hat \rho + m\rho\varpi^{-2}
\phi_a\,\deltaIRR\hat v^a
\label{e:FirstOrderEvenParityRhoEq}
\nonumber\\ 
&&\qquad-\nabla_a\left(\rho
P^a{}_b\deltaIRR \hat v^b\right)=\beta\, \deltaINR
\rho,\\ 
&&\!\!\!\!\!
\left[(\omega_N+m\Omega)\phi_a-2\varpi\Omega\nabla_a\varpi\right]
\deltaIRR\hat v^a+ m\,\deltaIRR \hat U\nonumber\\ 
&&\qquad= 
\phi_b\,\delta^{(1)}_\perp \hat f^b_{GR},\qquad\quad\\ 
&&\!\!\!\!\!
\left[(\omega_N+m\Omega)P^a{}_b-\frac2\varpi\Omega\nabla^a\varpi\phi_b\right]
\deltaIRR\hat v^b\nonumber\\ 
&&\qquad+ P^{ab}\nabla_b\deltaIRR \hat
U = P^a{}_b\,\delta^{(1)}_\perp \hat f^b_{GR}.
\label{e:FirstOrderOddParityVelocityEq} 
\eea
The general solution to the inhomogeneous system,
Eqs.~(\ref{e:FirstOrderEvenParityRhoEq})--(\ref{e:FirstOrderOddParityVelocityEq}),
for $\deltaIRR\hat\rho$, $\deltaIRR\hat U$, and $\deltaIRR\hat v^a$
consists of an arbitrary solution to the homogeneous equations
(obtained by setting $\beta\deltaINR\hat\rho =\delta^{(1)}_\perp
f_{GR}^a=0$) plus a particular solution.  These homogeneous equations
are identical to
Eqs.~(\ref{e:FirstOrderOddParityRhoEq})--(\ref{e:FirstOrderEvenParityVelocityEq}),
so their general solution is a multiple of the classical $r$-modes.
Because their $\phi$-parity is opposite to that of the classical
$r$-modes the effect of the homogeneous contributions
$\deltaIRR\hat\rho$, $\deltaIRR\hat U$, and $\deltaIRR\hat v^a$ is to
change the overall phase of the mode.  We choose (by appropriately
adjusting the time that we label $t=0$) to keep this phase unchanged,
and we can therefore, without loss of generality, set to zero the
homogeneous parts of the solutions to
Eqs.~(\ref{e:FirstOrderOddParityRhoEq})--(\ref{e:FirstOrderEvenParityVelocityEq}).
The inhomogeneous terms on the right sides of
Eqs.~(\ref{e:FirstOrderEvenParityRhoEq})--(\ref{e:FirstOrderOddParityVelocityEq}),
$\beta\deltaINR\hat\rho$ and $\delta^{(1)}_\perp \hat f_{GR}^a$, are all of
order $\beta$.  Thus the particular solution to
Eqs.~(\ref{e:FirstOrderEvenParityRhoEq})--(\ref{e:FirstOrderOddParityVelocityEq})
must also be of order $\beta$ as well.  It follows that the
radiation-reaction corrections to the first-order $r$-modes $\deltaIRR
Q$ are smaller than the classical $r$-modes $\deltaINR Q$ by terms of
order $\cal{O}(\beta/\omega)$.  To lowest-order in $\beta$, therefore,
the corrections to the first-order $r$-modes in
Eqs.~(\ref{e:deltaIrho})--(\ref{e:deltaIPhi}) simply change the
overall scale of the mode by the factor $e^{\beta t}$: $\deltaI Q =
\deltaINR Q\, e^{\beta t}$.

\subsection{Second-Order Perturbations}
\label{s:SecondOrderPerturbations}

The second-order perturbation equations are a sum of terms linear 
in $\deltaII Q$ and terms quadratic in $\deltaI Q$. For example, 
the second-order perturbation of the Euler equation, 
$\displaystyle\deltaII E^a = \left.\frac12\frac {d^2}{d\alpha^2}E^a\right|_{\alpha=0}$, 
includes the term $\deltaI v^b\nabla_b\deltaI v^a$, which serves as 
an effective source term for the second-order perturbations 
$\deltaII v^a$ and $\deltaII U$.  In the absence of gravitational 
radiation reaction, it follows that the
second-order Newtonian $r$-mode $\deltaIINR Q$ is a sum of 
terms of three kinds: a term with angular and temporal dependence 
$\cos(2\psi_N)$, where $\psi_N=m\phi+\omega_N t$, a term with dependence 
$\sin(2\psi_N)$,  
and a term that is time independent and axisymmetric.  
This time-independent axisymmetric part of the
velocity perturbation can be regarded as differential rotation.  
As we have emphasized in the Introduction, the second-order Newtonian $r$-modes are not determined uniquely:  
Given a particular solution $\deltaII_{NP} Q$ to the second-order Newtonian perturbation equations with perturbed velocity field $\deltaII_{NP} v^a$, 
there is a family of solutions $\deltaIINR Q$ with perturbed velocity field 
$\deltaIINR v^a = \deltaII_{NP} v^a + \deltaIINR\Omega(\varpi)\phi^a$, where  
$\deltaIINR\Omega(\varpi)$ is arbitrary.  
  This degeneracy is broken by
gravitational radiation reaction.  The presence of the
radiation-reaction force picks out a unique $\deltaII v^a$ that
displays the gravitational radiation driven growth of the second-order
$r$-modes: $\deltaII v^a\propto e^{2\beta t}$.

To find this differential rotation law, one must solve the second-order 
axisymmetric perturbation equations with radiation-reaction force for
the axisymmetric parts of the second-order $r$-modes. 
Denote the axisymmetric part of a perturbation $\delta Q$ by 
$\bigl\langle \delta Q \bigr\rangle$, and denote by $\delta^{(2)}\Omega$
the exponentially growing differential rotation of the unstable $r$-mode:
\be
 \deltaII\Omega \equiv \bigl\langle\deltaIINR v^\phi\bigr\rangle e^{2\beta t}
	= [\bigl\langle\deltaII_{NP} v^\phi\bigr\rangle 
	             + \deltaIINR\Omega(\varpi)] e^{2\beta t}. 
\label{e:Omega_decomp}\ee

Without solving the full system, however, one can obtain 
a formal expression for $\delta^{(2)}\Omega$ in terms of the 
known first-order perturbation together with 
other parts of the second-order axisymmetric perturbation.
As we will see in the next section, this expression is all that is 
needed to find $\delta^{(2)}\Omega$ to lowest nonvanishing 
order in $\Omega$: The other parts of the second-order perturbation 
give only higher-order contributions.  Finding this formal expression 
for $\delta^{(2)}\Omega$ and showing that it is 
unique are the goals of the present section.  

We now turn our attention to solving the perturbation equations for
the axisymmetric parts of the second-order $r$-modes.   The axisymmetric parts of the second-order
perturbations can be written in terms of their radiative and
nonradiative pieces:
\bsube
\bea
\bigl\langle\deltaII \rho \bigr\rangle&=& \Bigl(\bigl\langle \deltaIINR \rho\bigr\rangle
+\bigl\langle \deltaIIRR \rho\bigr\rangle\Bigr)e^{2\beta t},\\
\bigl\langle\deltaII v^a \bigr\rangle&=& 
\Bigl(\bigl\langle \deltaIINR v^a\bigr\rangle
+\bigl\langle \deltaIIRR v^a\bigr\rangle\Bigr)e^{2\beta t},\\
\bigl\langle\deltaII U \bigr\rangle&=& \Bigl(\bigl\langle \deltaIINR U\bigr\rangle
+\bigl\langle \deltaIIRR U\bigr\rangle\Bigr)e^{2\beta t},\\
\bigl\langle\deltaII \Phi \bigr\rangle&=& 
\Bigl(\bigl\langle \deltaIINR \Phi\bigr\rangle
+\bigl\langle \deltaIIRR \Phi\bigr\rangle\Bigr)e^{2\beta t},\\
\bigl\langle\deltaII f^a_{GR} \bigr\rangle&=& 
\bigl\langle \deltaIIRR f^a_{GR}\bigr\rangle e^{2\beta t}.
\eea\esube
These quantities are determined by the second-order axisymmetric parts
of the perturbed stellar evolution equations:
\bea
&&
\!\!\!\!\!\!
2\beta\bigl\langle \deltaII v^a\bigr\rangle 
+ 2\Omega\bigl\langle\deltaII v^b\bigr\rangle\nabla_b\phi^a
+\nabla^a\bigl\langle\deltaII U\bigr\rangle
\nonumber\\
&&\qquad\qquad\qquad = \bigl\langle \deltaII\! f^a_{GR}\bigr\rangle
-\bigl\langle\deltaI v^b\,\nabla_b\deltaI v^a\bigr\rangle,
\label{e:PerturbedEulerII}\qquad\\
&&
\!\!\!\!\!\!
2\beta\bigl\langle\deltaII \rho\bigr\rangle + \nabla_a\Bigl[
\rho \bigl\langle\deltaII v^a\bigr\rangle + \bigl\langle \deltaI \rho\, \deltaI v^a\bigr\rangle
\Bigr] = 0,
\label{e:PerturbedMassConsII}\\
&&
\!\!\!\!\!\!
\nabla^2 \bigl\langle\deltaII\Phi\bigr\rangle = 4\pi \bigl\langle\deltaII
\rho\bigr\rangle.
\label{e:PerturbedPoissonII}\eea

The uniqueness of the second-order differential rotation $\deltaII\Omega$ 
can be seen as follows. Let $\langle\deltaII Q\rangle$ and 
$\langle\deltaII \widetilde Q\rangle$ be two solutions to the 
second-order perturbation equations, Eqs.~(\ref{e:PerturbedEulerII}), 
(\ref{e:PerturbedMassConsII}), and (\ref{e:PerturbedPoissonII}),
associated with the same time-dependence $e^{2\beta t}$ and with the same 
first-order solution $\deltaI Q$.  The difference 
$\langle\deltaII Q\rangle - \langle\deltaII \widetilde Q\rangle$
of the two solutions then satisfies the {\it linearized} Poisson equation and 
the {\it linearized} Euler and mass conservation equations obtained by 
setting to zero the terms involving $\deltaI v^a$ and $\deltaII f^a_{GR}$ 
in Eqs.~(\ref{e:PerturbedEulerII}) and (\ref{e:PerturbedMassConsII}). 
That is, $(\langle\deltaII Q\rangle - \langle\deltaII \widetilde Q\rangle)e^{2\beta t}$ 
is an axisymmetric solution to the first-order Newtonian perturbation equations. 
But the Newtonian star has no such solution, no mode with growth rate $2\beta$.
Thus $(\langle\deltaII Q\rangle - \langle\deltaII \widetilde Q\rangle)e^{2\beta t}=0$,
implying that $\deltaII\Omega$ is unique. 
(Note, however, that the decomposition (\ref{e:Omega_decomp}) is not unique: 
The arbitrariness in the differential rotation of the Newtonian $r$-mode 
means that one is free to add to $\bigl\langle\deltaII_{NP}v^\phi\bigr\rangle$ 
an arbitrary function $f(\varpi)$ if one simultaneously changes $\deltaIINR\Omega(\varpi)$ to $\deltaIINR\Omega(\varpi) - f(\varpi)$.)   

We now obtain equations for $\deltaIINR Q$ and $\deltaIIRR Q$.   
Keeping terms to first order in $\beta$, the terms quadratic in
first-order perturbed quantities that appear in
Eqs.~(\ref{e:PerturbedEulerII}) and (\ref{e:PerturbedMassConsII}) have
the forms,
\bea
\bigl\langle\deltaI v^b\nabla_b\deltaI v^a\bigr\rangle &=&
\left(\bigl\langle\deltaINR v^b\nabla_b
\deltaINR v^a\bigr\rangle\right.\nonumber\\
&&\qquad\qquad\left.
+\beta\,\bigl\langle\deltaIIRR V^a\bigr\rangle\right) e^{2\beta t},\\
\bigl\langle\deltaI \rho\, \deltaI v^a \bigr\rangle &=&
\left(\bigl\langle\deltaINR \rho\, 
\deltaINR v^a \bigr\rangle 
+\beta\, \bigl\langle\deltaIIRR W^a\bigr\rangle\right) e^{2\beta t},\nonumber\\
\eea
where
\bea
\beta \bigl\langle\deltaIIRR V^a\bigr\rangle&=&
\bigl\langle\deltaIRR v^b\nabla_b
\deltaINR v^a\bigr\rangle
+\bigl\langle\deltaINR v^b\nabla_b\deltaIRR v^a\bigr\rangle,
\label{e:Vdef}\qquad\\
\beta \bigl\langle\deltaIIRR W^a\bigr\rangle&=& 
\bigl\langle\deltaIRR \rho\, 
\deltaINR v^a \bigr\rangle + \bigl\langle\deltaINR \rho\, 
\deltaIRR v^a \bigr\rangle .\label{e:Wdef}
\eea

The nonradiative parts $\langle\deltaIINR Q\rangle$ of the perturbations 
are determined, up to a perturbation that adds differential rotation 
$\deltaIINR\Omega(\varpi)$, 
by the axisymmetric parts of the Newtonian Euler and 
mass-conservation equations:
\bea
&&
\!\!\!\!
2\Omega\bigl\langle\deltaIINR v^b\bigr\rangle\nabla_b\phi^a
+\nabla^a\bigl\langle\deltaIINR U\bigr\rangle=
-\bigl\langle\deltaINR v^b\,\nabla_b\deltaINR v^a\bigr\rangle,
\label{e:NRPerturbedEulerII}\nonumber\\\\
&&
\qquad\quad
\nabla_a\Bigl[
\rho\, \bigl\langle\deltaIINR v^a\bigr\rangle + \bigl\langle \deltaINR\! \rho\,\, 
\deltaINR v^a\bigr\rangle
\Bigr] = 0.
\label{e:NRPerturbedMassConsII}
\eea
Given a particular solution $\deltaII_{NP} Q$ to these equations, we 
want to find the remaining contribution $\deltaIINR\Omega(\varpi)$ 
to the differential rotation of Eq.~(\ref{e:Omega_decomp})
that is picked out by the radiation-reaction.     
 
We define the radiative part of the perturbation,
$\bigl\langle\deltaIIRR Q\bigr\rangle$,
by requiring that it be created 
entirely by the radiation reaction forces; $\bigl\langle\deltaIIRR Q\bigr\rangle$ is therefore proportional 
to the radiation reaction rate $\beta$.  When $\langle\deltaIINR Q\rangle$ satisfies 
the Newtonian equations (\ref{e:NRPerturbedEulerII}) and (\ref{e:NRPerturbedMassConsII}), the axisymmetric parts of the 
full perturbed Euler and mass-conservation equations with radiation-reaction 
have at ${\cal O}(\beta)$ the form
\bea
&&
2\beta\bigl\langle\deltaIINR v^a\bigr\rangle
+ 2\Omega \bigl\langle \deltaIIRR v^b\bigr\rangle
\nabla_b\phi^a + \nabla^a\bigl\langle\deltaIIRR U\bigr\rangle
\nonumber\\
&&\qquad\qquad\qquad\qquad =\bigl\langle \deltaIIRR\! f_{GR}^{\,a}\bigr\rangle
-\beta\, \bigl\langle\deltaIIRR V^a\bigr\rangle,\\
&&
\label{e:SecondOrderPerturbedEuler}\nonumber\\
&&\nabla_a\Bigl(\rho\,\bigl\langle\deltaIIRR v^a \bigr\rangle\Bigr)
	= -2\beta\bigl\langle \deltaIINR \rho\bigr\rangle-\beta\,\nabla_a
\bigl\langle\deltaIIRR W^a\bigr\rangle.\qquad
\label{e:SecondOrderPerturbedMassConservation}
\eea 

To find an expression for $\deltaIINR\Omega(\varpi)$, we first write  
$\bigl\langle\deltaIINR v^a\bigr\rangle$ as 
$\bigl\langle\deltaII_{NP} v^a\bigr\rangle + \deltaIINR\Omega(\varpi)\phi^a$ and move the term involving 
$\bigl\langle\deltaII_{NP} v^a\bigr\rangle$ to the right side of 
Eq.~(\ref{e:SecondOrderPerturbedEuler}): 
\bea
2\beta\deltaIINR\Omega(\varpi)\phi^a
	+2\Omega \bigl\langle \deltaIIRR v^b\bigr\rangle\nabla_b\phi^a
	&+& \nabla^a\bigl\langle\deltaIIRR U\bigr\rangle \nonumber\\ 
&=&\beta \bigl\langle \deltaIIRR F^a\bigr\rangle,
\label{e:SecondOrderPerturbedEuler1}\qquad
\eea
where
\bea
\beta \bigl\langle \deltaIIRR F^a\bigr\rangle
=\bigl\langle \deltaIIRR\! f_{GR}^{\,a}\bigr\rangle
- 2\beta \bigl\langle\deltaII_{NP} v^a \bigr\rangle
-\beta\, \bigl\langle\deltaIIRR V^a\bigr\rangle.\quad
\label{e:Fdef}
\eea

We next write the components of the axisymmetric part
of the second-order perturbed Euler
equation, Eq.~(\ref{e:SecondOrderPerturbedEuler1}), in cylindrical coordinates:
\bsube
\bea
2\beta\varpi\deltaIINR\Omega(\varpi) 
 &+&2\Omega\bigl\langle \deltaIIRR v^\varpi\bigr\rangle
=\beta\varpi\bigl\langle \deltaIIRR F^{\,\phi}\bigr\rangle,
\label{e:SecondOrderEulerVarpi}
\qquad\\
-2\Omega\varpi\bigl\langle\deltaIIRR v^\phi\bigr\rangle 
 &=& -\partial_\varpi \bigl\langle\deltaIIRR U\bigr\rangle
	+\beta\bigl\langle \deltaIIRR F^{\varpi}\bigr\rangle,
\label{e:SecondOrderEulerPhi}\\
0&=& 
-\partial_z\bigl\langle\deltaIIRR U\bigr\rangle+\beta\bigl\langle \deltaIIRR F^{z}\bigr\rangle.
\qquad\label{e:SecondOrderEulerZ}
\eea\label{e:SecondOrderEuler}\esube
Using Eq.~(\ref{e:SecondOrderEulerVarpi}) to determine $\bigl\langle
\deltaIIRR v^\varpi\bigr\rangle$, the axisymmetric part of
the second-order mass conservation
Eq.~(\ref{e:SecondOrderPerturbedMassConservation}) can be written as
\bea 
&&\frac{\beta}{2\Omega\varpi}
\partial_\varpi\left[\rho\varpi^2\left(\bigl\langle \deltaIIRR F^{\,\phi}\bigr\rangle
  -2\deltaIINR\Omega(\varpi)\right)\right]\nonumber\\ &&\quad
+ \partial_z\left[\rho\bigl\langle \deltaIIRR
  v^z\bigr\rangle\right] = 
-2\beta\bigl\langle \deltaIINR \rho\bigr\rangle-\beta\,\nabla_a
\bigl\langle\deltaIIRR W^a\bigr\rangle.
\nonumber\\
\label{e:SecondOrderMassCons}  
\eea

The star's surface is defined as the $p=0$ surface.  Because
$\delta^{(2)}\rho$ is a derivative evaluated at $\alpha=0$, it has
support on the unperturbed star. While the density perturbation
$\deltaII\rho$ is not finite for some equations of state at the
surface of the star, it is integrable in the sense that $\deltaII\int
\rho\,dz$ is finite, as one would expect from the integrability of
the mass-conservation condition in
Eq.~(\ref{e:MassConservationIntegral}).  In particular, for polytropes
with fractional polytropic index $0<n<2$, $\deltaII\rho$ diverges at
$z=z_S$, but, as we show in Appendix \ref{s:surface}, $\deltaII\int
\rho\,dz$ is finite.   Here we  
denote by $z_S(\varpi)$ the value of $z$ (the Cartesian coordinate
axis parallel to the rotation axis) at the surface of the
unperturbed star.  

We now multiply the second-order mass conservation equation,
Eq.~(\ref{e:SecondOrderMassCons}), by $2\varpi\Omega/\beta$ and
integrate with respect to $z$ over the support of the star.  It will
be convenient to extend the domain of integration to extend slightly
beyond the surface of the unperturbed star. Because each integrand has
support on the unperturbed star, we simply take the integrals to
extend from $-\infty$ to $\infty$ instead of
$-z_S$ to $z_S$.  We then have
\bea
0&=&4\varpi \Omega\int_{-\infty}^{\infty} dz\bigl\langle \deltaIINR\rho\bigr\rangle\nonumber\\
&&+\int_{-\infty}^{\infty} dz\partial_\varpi\left[\rho\varpi^2
\left(\bigl\langle \deltaIIRR F^{\,\phi}\bigr\rangle
  -2\deltaIINR\Omega(\varpi) \right)
\right]\nonumber\\
&&+2\varpi\Omega \int_{-\infty}^{\infty} dz \nabla_a
\bigl\langle\deltaIIRR W^a\bigr\rangle.
\label{e:Integral1}
\eea 
The second integral on the right side of Eq.~(\ref{e:Integral1}) can
be rewritten as
\bea 
&&\int_{-\infty}^{\infty}
dz\partial_\varpi\left[\rho\varpi^2 \left(\bigl\langle \deltaIIRR
  F^{\,\phi}\bigr\rangle -2\deltaIINR \Omega(\varpi)
  \right)\right]=\nonumber\\ 
&&\qquad\qquad\partial_\varpi\int_{-\infty}^{\infty}
dz\rho\varpi^2 \left(\bigl\langle \deltaIIRR F^{\,\phi}\bigr\rangle
-2\deltaIINR \Omega(\varpi)\right).\qquad 
\label{e:Integral2}
\eea 

The expression in Eq.~(\ref{e:Integral1}) can
then be integrated from $\varpi=0$ to $\varpi$, using
Eq.~(\ref{e:Integral2}), to obtain an expression for
$\deltaIINR\Omega(\varpi) $:
\bea
&&\!\!\!\!\!\!
2\varpi^2 \deltaIINR \Omega(\varpi) \int_{-\infty}^{\infty} dz\, \rho =
\varpi^2 \int_{-\infty}^{\infty} dz\, \rho\, 
\bigl\langle \deltaIIRR F^{\,\phi}\bigr\rangle\nonumber\\
&&\quad+4\Omega \int_0^\varpi d\varpi' \varpi' 
\int_{-\infty}^{\infty} dz\bigl\langle \deltaIINR\rho\bigr\rangle
\nonumber\\
&&\quad+2\Omega\int_0^\varpi d\varpi' \varpi' \int_{-\infty}^{\infty} dz
\,\nabla_a\bigl\langle\deltaIIRR W^a\bigr\rangle.
\label{e:Integral3}
\eea 
Because of the axisymmetry of its integrand, the third term on the
right side of Eq.~(\ref{e:Integral3}) is 
the volume integral of a divergence. The boundary of the
three-dimensional region of integration has two parts: One is 
outside the surface of the star, where $\deltaIIRR W^a$ vanishes; the
second is the cylinder at constant $\varpi$ from $-z_S$ to $z_S$, with
outward normal $\nabla_a\varpi$ and element of area $\varpi d\phi
dz$. The term is then given by
\begin{align}
{\displaystyle \int}_0^\varpi d\varpi' \varpi' \int_{-\infty}^{\infty} dz
\,\nabla_a&\bigl\langle\deltaIIRR W^a\bigr\rangle
= \varpi \int_{-\infty}^{\infty} dz&\bigl\langle\deltaIIRR W^\varpi\bigr\rangle.\qquad
\end{align}
With this simplification, Eq.~(\ref{e:Integral3})
can be written in the form:
\bea
&&\!\!\!\!\!\!
2\varpi^2 \deltaIINR \Omega(\varpi) \int_{-\infty}^{\infty} dz\, \rho =
\varpi^2 \int_{-\infty}^{\infty} dz\, \rho\, 
\bigl\langle \deltaIIRR F^{\,\phi}\bigr\rangle\nonumber\\
&&\quad+4\Omega \int_0^\varpi d\varpi' \varpi' 
\int_{-\infty}^{\infty} dz\bigl\langle \deltaIINR\rho\bigr\rangle\nonumber\\
&&
\quad+2\varpi\Omega\int_{-\infty}^{\infty} dz\bigl\langle\deltaIIRR W^\varpi\bigr\rangle.
\label{e:Integral4}
\eea

This provides a formal expression for $\deltaIINR \Omega(\varpi)$ in terms of the
first-order perturbations that comprise $\bigl\langle \deltaIIRR
F^{\,\phi}\bigr\rangle$ and $\bigl\langle\deltaIIRR W^\varpi\bigr\rangle$ 
and the second-order perturbation $\bigl\langle \deltaIINR\rho\bigr\rangle$.
\footnote{
As mentioned above, Appendix \ref{s:surface} shows that 
assuming smoothness of the displacement of the surface as a 
function of $\alpha$ and $\vec x$ implies integrability of 
$\deltaIINR\rho$. A simpler way to see that the right side of 
Eq.~(63) is finite is to note that smoothness of the displacement of 
the surface implies one-sided differentiability of $\deltaII\vec v$ 
at the surface. 
The perturbed mass conservation equation,   
Eq.~(\ref{e:SecondOrderPerturbedMassConservation}), 
then implies that the combination 
$ 2 \langle\deltaIINR\rho\rangle + \nabla_a \langle\deltaIIRR W^a\rangle$ 
is integrable.   
This is enough to imply that the expression in Eq.~(\ref{e:Integral4}) for 
$\deltaIINR \Omega(\varpi)$ is finite. 
}
  
Together with $\bigl\langle\deltaII_{NP} v^\phi\bigr\rangle $, 
it determines the differential rotation of the unstable $r$-mode.

We conclude this section with a discussion of two simplifications in 
evaluating $\deltaIINR \Omega(\varpi)$, one from the 
fact that we work to first order in the growth rate $\beta$, the 
second from the slow-rotation approximation of the next section.      
The first is a simplification of the expression Eq.~(\ref{e:Integral4})
for  the radiation-reaction force.  The integrand of the first term in 
Eq.~(\ref{e:Integral4}), $\rho\bigl\langle \deltaIIRR F^{\,\phi}\bigr\rangle$, 
is given by the $\phi$-component of Eq.~(\ref{e:Fdef}): 
\be \beta \bigl\langle
\deltaIIRR F^{\,\phi}\bigr\rangle = \bigl\langle \deltaIIRR\!
f_{GR}^{\,\phi}\bigr\rangle - 2\beta \bigl\langle\deltaIINR v^\phi
\bigr\rangle -\beta\, \bigl\langle\deltaIIRR V^\phi\bigr\rangle.
\label{e:betaF}
\ee 
To evaluate $\bigl\langle \deltaIIRR\! f_{GR}^{\,\phi}\bigr\rangle$,
we must find the axisymmetric, second-order, part of the expression
for $\delta\vec f_{GR}$ on the right side of
Eq.~(\ref{e:PerturbedGRForceExact}).  Recall that the axisymmetric
parts of any second-order quantity have time dependence $e^{2\beta
  t}$.  The first three terms in the bracketed expression in
Eq.~(\ref{e:PerturbedGRForceExact}) involve high-order time derivatives of
$\delta^{(2)} I^{\ell 0}$ or $\delta^{(2)} S^{\ell 0}$, and are
therefore proportional to high powers of $\beta$ and can be neglected.  
We are left with only the fourth term,
\bea
\big\langle\deltaIIRR \vec f_{GR}\bigr\rangle &=& \frac{(-1)^\ell
  N_\ell}{8\pi\sqrt{\ell}}\nonumber\\ &&\times
\Re\left\langle\deltaINR \vec v \times \vec\nabla(r^\ell Y^{\ell\ell})
\frac{d^{\,2\ell+1}\deltaINR
  S^{\ell\ell}}{dt^{\,2\ell+1}}\right\rangle.\nonumber\\
\label{e:fGR}
\eea 
 
The second simplification involves the quantities 
$\bigl\langle\deltaIIRR V^a\bigr\rangle$ and
$\bigl\langle\deltaIIRR W^a\bigr\rangle$ that appear 
in Eq.~(\ref{e:Integral4}). They are defined in
Eqs.~(\ref{e:Vdef}) and (\ref{e:Wdef}).  Using the general expressions
for the first order perturbations given in
Eqs.~(\ref{e:deltaIrho})--(\ref{e:deltaIPhi}), we can express these
quantities in terms of the first order perturbations:
\begin{eqnarray}
\bigl\langle\beta\deltaIIRR
W^a\bigr\rangle&=&\half P^a{}_b\left(\deltaIRR\hat\rho\deltaINR\hat v^b
+\deltaINR\hat\rho\deltaIRR\hat v^b\right),\quad\\
\bigl\langle\beta\deltaIIRR V^a\bigr\rangle&=&
\half\varpi^{-2}\phi^a\left[\deltaIRR\hat v^k
\nabla_k\left(\deltaINR\hat v^b\phi_b\right)
\right.\nonumber\\
&&\qquad\qquad
\left.+\deltaINR\hat v^k\nabla_k\left(\deltaIRR\hat v^b\phi_b\right)\right].
\end{eqnarray}
As we will see in the following section, these terms and the term 
involving $\deltaIINR\rho$ in Eq.~(\ref{e:Integral4}) are higher order
in $\Omega$ than the first two terms of Eq.~(\ref{e:betaF}) 
and can therefore be neglected when evaluating
$\deltaIINR\Omega(\varpi)$ for slowly rotating stars using
Eq.~(\ref{e:Integral4}).  
This fact is essential, because $\deltaIINR\rho$ itself depends 
on $\deltaIINR\Omega$.  

This discussion has been somewhat abstract but quite general.  Apart
from assuming the integrability of the perturbed density so that mass
conservation, Eq.~(\ref{e:MassConservationIntegral}), can be enforced,
no assumption has been made up to this point about the particular
equation of state of the matter in these stellar models, nor has any
assumption been made about the magnitude of the angular velocity of
the star.  In order to proceed further, however, we will need to
assume that the stellar model is slowly rotating in a suitable sense.
To find an explicit solution for $\deltaIINR\Omega(\varpi)$, we will also need
to make some choice for the equation of state for the stellar matter.
The slow rotation assumption and its implications are discussed in
Sec.~\ref{s:SlowRotation}, while the complete solution for
$\deltaII\Omega$, the second-order $r$-mode angular velocity that is
driven by gravitational radiation reaction, is determined in
Sec.~\ref{s:PolytropicStellarModels} for the case stars of composed of
matter with a range of polytropic equations of state.

\section{Slow Rotation Expansion}
\label{s:SlowRotation}

We consider the one-parameter families of stars $Q=Q(\Omega)$ composed
of matter with a fixed equation of state, and having masses that are
independent of the angular velocity: $M(\Omega)=M_0$.  The structures
of slowly rotating stellar models in these families are conveniently
written as expansions in the dimensionless angular velocity,
\be
\widetilde\Omega =\frac{\Omega}{\Omega_0},
\ee
where $\Omega_0=\sqrt{M_0/R^3}$, and $M_0$ is the mass and $R$ the radius
of the non-rotating star in the sequence.  The slow rotation expansion of these stellar models
is denoted,
\be
Q = \sum_{n=0} Q_n \widetilde\Omega^n = Q_0 + Q_1 \widetilde\Omega + Q_2 \Omega^2
+ {\cal O}(\widetilde\Omega^3). 
\label{e:Qn}\ee
For equilibrium rotating stars these expansions of the basic fluid
variables have the forms:
\bea
\rho&=&\rho_0+\rho_2\,\widetilde\Omega^2+{\cal O}(\Omega^4),\\
v^a&=&\Omega\,\phi^a,\\
p&=&p_0+p_2\,\widetilde\Omega^2+{\cal O}(\Omega^4),\\
\Phi&=&\Phi_0+\Phi_2\widetilde\Omega^2+{\cal O}(\Omega^4).
\eea
We will represent the perturbations of these stellar models $\delta Q$
as dual expansions in the mode amplitude $\alpha$ and the angular
velocity parameter $\widetilde\Omega$:
\bea
\delta Q = \sum_{n,k} \alpha^n\, \widetilde\Omega^k\, \deltaN Q_k.
\eea

Our main goal here is to determine to lowest-order in angular velocity
the axisymmetric part of the second-order perturbations of the
$r$-mode angular velocity field $\bigl\langle \deltaIIRR
v^\phi\bigr\rangle $ that is driven by the gravitational-radiation
instability.  Doing this requires the explicit slow-rotation forms of
the first and the second-order perturbations.  These slow-rotation
expansions are described in the remainder of this section.  

\subsection{First Order Perturbations}  
\label{s:slowfirstorder}

The effect of the first-order gravitational radiation-reaction force 
$\deltaI \vec f_{GR}$ on the structure
of the classical $r$-mode (beyond its overall effect
on its amplitude) was first studied (for $\ell=2$) by 
Dias and S\'a \cite{Sa2005b}.  We agree with the results they obtain but will 
need to clarify their meaning.  We also extend the calculation to general 
values of $\ell$.  

To first order in mode amplitude $\alpha$ and lowest non-trivial
order in angular velocity $\tilde \Omega$, the classical $r$-modes
with the $\phi$-parity described in Sec.~\ref{s:FirstOrderPerturbations}
can be written the form
\bea
\deltaINR p_1&=&\deltaINR \rho_1=\deltaINR \Phi_1=0,
\label{e:ClassicalRModeDensity}
\\
\deltaINR \vec v_1&=&\Im\left[\frac{R\Omega_0}{\ell} 
\left(\frac{r}{R}\right)^{\ell}\!\!\vec r\times
\vec\nabla\left(\sin^\ell\theta e^{i\ell\phi+i\omega t}\right)\right],\nonumber\\ 
\label{e:ClassicalRModeVelocity}
\eea
where $\Im(Z)$ is the imaginary part of a quantity $Z$.  
An equivalent expression
for the classical $r$-mode velocity in terms of vector spherical harmonics is 
\bea \deltaINR \vec v_1 &=& \Im\left(A_\ell r^\ell \vec Y^{\ell \ell}_B
e^{i\omega t}\right),\\ &=&-\frac{iA_\ell r^\ell}{2}\left[\vec
Y^{\ell\ell}_Be^{i\omega t}-(-1)^\ell \vec Y^{\ell-\ell}_Be^{-i\omega
  t}\right], 
\eea 
where $A_\ell$ is given by
\begin{eqnarray}
A_\ell&=& (-1)^\ell 2^\ell (\ell-1)! \sqrt{\frac{4\pi\ell(\ell+1)}{(2\ell+1)!}}
R^{-\ell+1}\Omega_0.
\end{eqnarray}
The frequencies of these classical $r$-modes have the form
\bea
\omega_N
&=&-\frac{(\ell-1)(\ell+2)}{\ell+1}\Omega +{\cal O}(\Omega^3).
\label{e:RModeFrequency}
\eea
At this order in $\Omega$, the $r$-modes do not affect the fluid variables  
$\delta \rho$ and $\delta p$, which are ${\cal O}(\Omega^2)$. Because of this,
the $r$-mode velocity field at order $\Omega$  
does not depend on the equation of state.

Four features of the gravitational radiation-reaction force are
important in determining the way it alters each $r$-mode: {\it a)} The
$\phi$-parity of $\deltaI \vec f_{GR} $, as shown in the last section,
is opposite to that of the classical mode; {\it b)} its magnitude, as
shown below, is dominated by the current current multipole
$S^{\ell\ell}$; {\it c)} it can be decomposed in the manner \be
\deltaI \vec f_{GR} = \beta \deltaINR\vec v + \delta^{(1)}_\perp \vec f_{GR},
\label{e:fdecompose}\ee 
where the two terms in the decomposition are orthogonal with respect
to a density-weighted inner product, $\int \sqrt{g}\, d^{\,3}x\,
\rho_0\, \deltaINR\vec v\,\cdot\, \delta^{(1)}_\perp \vec f_{GR} =0$;
and {\it d)} as we show below, $\delta^{(1)}_\perp \vec f_{GR}$ is a
gradient, $\delta^{(1)}_\perp \vec f_{GR} = \vec\na
\delta^{(1)}_\perp{\cal F}$.

It is straightforward to evaluate the multipole moments of the $r$-modes
using Eqs.~(\ref{e:PerturbedMassMultipole}) and
(\ref{e:PerturbedCurrentMultipole}) and the expressions for the
classical $r$-modes from Eqs.~(\ref{e:ClassicalRModeDensity}) and
(\ref{e:ClassicalRModeVelocity}).  The expressions for the
non-vanishing multipole moments of the $r$-modes can be written in the
form
\bea
\!\!\!\!\!\!
\deltaINR S^{\ell\ell} &=& (-1)^\ell \deltaINR S^{*\ell-\ell}\nonumber\\
&=&-i\frac{A_\ell N_\ell e^{i\omega t} }{\sqrt{\ell+1}}
\int_0^R r^{2\ell+2}\rho_0\, dr.
\label{e:deltaS}\eea
Inserting these expressions into the formula for the
gravitational radiation-reaction force,
Eq.~(\ref{e:PerturbedGRForceExact}), we find
\bea
\!\!\!\!\!\!
\deltaINR \vec f_{GR} &=& \frac{(-1)^\ell N_\ell}{8\pi}
\Re\left\{\left[\frac{i\omega}{\sqrt{\ell+1}}
r^\ell \vec Y^{\ell\ell}_B\right.\right. \nonumber\\
&&+\left.\left. \frac{\Omega}{\sqrt{\ell}}
\vec\phi
\times\vec\nabla(r^\ell Y^{\ell\ell})\right]
\frac{d^{2\ell+1}\delta S^{\ell\ell}}{dt^{2\ell+1}}\right\}.
\label{e:FirstOrderRModeGRForce}
\eea
This expression can be rewritten as a linear combination
of $r^\ell\vec Y^{\ell\ell}_B$ and $\vec\nabla(r^\ell Y^{\ell\ell})$
using the identity
\bea
\vec\phi\times\vec\nabla(r^\ell Y^{\ell\ell}) =
i\sqrt{\ell(\ell+1)}r^\ell\vec Y^{\ell\ell}_B - z \vec\nabla(r^\ell Y^{\ell\ell}).
\eea
The resulting expression for $\deltaINR \vec f_{GR}$ can
therefore be written in the following way:
\bea
\deltaI \vec f_{GR} &=& \beta \deltaINR\vec v + \delta^{(1)}_\perp \vec f_{GR},
\label{e:FirstOrderGRForce}
\eea
where $\beta$ is given by
\bea
\beta = \frac{N^2_\ell \omega^{2\ell+2}}{4\pi(\ell^2-1)(\ell+2)}
\int_0^R r^{2\ell+2}\rho_0\,dr,
\label{e:BetaDef}
\eea
and where $\delta^{(1)}_\perp \vec f_{GR}$ is defined by
\bea
&&\!\!\!\!\!
\delta^{(1)}_\perp \vec f_{GR}=
-\frac{N^2_\ell \omega^{2\ell+1}\Omega}{8\pi}
\int_0^R r^{2\ell+2}\rho\,dr\nonumber\\
&&\qquad\qquad\times
\left\{\frac{\deltaINR \vec v}{\ell+1}
+\frac{\Re\left[z A_\ell\vec\nabla(r^\ell Y^{\ell\ell})e^{i\omega t}\right]}
{\sqrt{\ell(\ell+1)}}\right\}
.\qquad 
\label{e:F_R_GR_Def}
\eea
This expression for $\delta^{(1)}_\perp \vec f_{GR}$ can be rewritten as 
a gradient, 
\bea
\delta^{(1)}_\perp \vec f_{GR}&=&\Im\left\{
i\beta A_\ell\frac{\sqrt{\ell(\ell+1)}}2\vec\nabla\left[
r^{\ell+1}\cos\theta\, Y^{\ell\ell}e^{i\omega t}\right]\right\}\nonumber\\
&=:&\vec\nabla \delta^{(1)}_\perp{\cal F}.
\label{e:Fgradient}
\eea
Eqs.~(\ref{e:FirstOrderGRForce}) and (\ref{e:Fgradient}) give the 
decomposition of Eq.~(\ref{e:fdecompose}),    
and the orthogonality of the two parts,  
\be 
\int\rho_0 \deltaINR \vec v \cdot \delta^{(1)}_\perp \vec f_{GR}
\,\sqrt{g}\,d^{\,3}x = 0, \ee 
is implied by the relation
\bea
&&
\displaystyle \int \epsilon^{abc} \nabla_a(\cos\theta\,
Y^{\ell\ell}) \nabla_b r \nabla_c\bar
Y^{\ell\ell}\sqrt g\,d^{\,2}\,x\nonumber\\ 
&&\qquad = -\int 
\epsilon^{abc} \cos\theta\, Y^{\ell\ell}\nabla_b r\nabla_a
\nabla_c\bar Y^{\ell\ell} \sqrt g\,d^{\,2}\,x= 0,\nonumber\\ 
\eea
where $\sqrt{g}\,d^{\,2}\,x$ is the volume element on the sphere:
$\sqrt{g}\,d^{\,2}\,x\equiv -r^2d\cos\theta\, d\phi$.  At this order
in $\Omega$, the density $\rho_0$ plays no role in the orthogonality,
but it is with respect to the density-weighted inner product that the
operators appearing in the perturbed Euler equation are formally
self-adjoint.
 
It follows that $\delta^{(1)}_\perp \vec f_{GR}$ is the part of the 
gravitational radiation-reaction force that does not contribute
directly to the exponential growth of the classical $r$-mode
instability and that the coefficient $\beta$ is the
growth rate of the gravitational radiation driven instability in the
$r$-modes.  Substituting into Eq.~(\ref{e:BetaDef}) the expressions for
$N_\ell$ from Eq.~(\ref{e:NlDef}) and the $r$-mode frequency $\omega$
from Eq.~(\ref{e:RModeFrequency}) gives
\bea
\!\!\!\!
\beta= \frac{32\pi\Omega^{2\ell+2}(\ell-1)^{2\ell}}{[(2\ell+1)!!]^2}\!\!
\left(\frac{\ell+2}{\ell+1}\right)^{2\ell+2}\!\!
\int_0^R r^{2\ell+2}\rho_0\,dr,
\label{e:beta}\eea
which agrees with the expression for the gravitational radiation
growth rate of the $r$-mode instability given in Lindblom,
Owen and Morsink~\cite{Lindblom98b}. 

These expressions for the slow rotation limits of the
radiation-reaction force confirm the general expressions,
e.g. Eq.~(\ref{e:GRForceParityEq}), used in our discussion of the
general properties of the first-order $r$-modes in
Sec.~\ref{s:FirstOrderPerturbations}.  It follows from that discussion
that the general form of the first-order $r$-mode
velocity, to lowest order in
the angular velocity of the star, is given by
\bea
\deltaI \vec v = \tilde \Omega\, \deltaINR\! \vec v_1\, e^{\beta t}.
\eea

To evaluate $\deltaIINR\Omega$ using Eq.~(\ref{e:Integral4}), we
need to determine $\deltaIRR\rho$ and $\deltaIRR \vec v$, or at least
to show that they are negligibly small compared to other terms in the
equation.  We show in the heuristic argument below that
$\deltaIRR\rho={\cal O}(\beta\Omega) $ and $\delta_R^{(1)} \vec v =
{\cal O}(\beta\Omega^2)$, which will allow us to neglect them in our
slow rotation expansion.  A more precise version of the argument is 
given in Appendix~\ref{s:ordering_appendix}.  
The fact that $\deltaIRR \vec v$ is
higher-order in $\Omega$ than $\deltaIRR\rho$ is the reverse of their
relation in the classical $r$-modes.  This reversal depends on the
appearance of the gradient $\vec \nabla \delta^{(1)}_\perp{\cal F}$ in the
decomposition of the gravitational radiation-reaction force $\deltaIRR
\vec f_{GR}$.

The equations that determine $\deltaIRR Q$,
Eqs.~(\ref{e:FirstOrderEvenParityRhoEq})--(\ref{e:FirstOrderOddParityVelocityEq}),
can be written more compactly as
\bea
(\omega_N+\ell\,\Omega)\,\deltaIRR\hat \rho +\vec\nabla\cdot\left(\rho\deltaIRR
\vec{\hat v}\right) &=&\beta\, \deltaINR\rho,
\label{e:d1continuity}\\
(\omega_N+\ell\,\Omega)\deltaIRR\vec{\hat v} +2\Omega\deltaIRR\vec{\hat v}\cdot 
\nabla \vec\phi 
	&& \nonumber\\
=  -\vec\nabla(\deltaIRR \hat U\!\!&-&\!\!\delta^{(1)}_\perp{\cal F}).  
\label{e:d1Euler} 
\eea
The value of $\delta_R^{(1)} \vec{\hat v}$ is fixed by 
the curl of the perturbed Euler equation, (\ref{e:d1Euler}):
\be
 \vec\na\times\left[(\omega_N+\ell\,\Omega)
\deltaIRR \vec{\hat v} +2\Omega\deltaIRR \vec{\hat v}\cdot \na\vec \phi\right]=0, 
\ee
which involves only $\delta_R^{(1)}\vec{\hat v}$. Its two independent
components give two relations for the three components of
$\delta_R^{(1)}\vec{\hat v}$, in which all coefficients are ${\cal
  O}(\Omega)$. All components of $\delta_R^{(1)}\vec{\hat v}$ are
therefore of the same order in $\Omega$.  Similarly, the two relations
among $\delta_R^{(1)}U$, $\delta_R^{(1)}\Phi$, and
$\delta_R^{(1)}\rho$ given by the equation of state and the Poisson
equation imply that $\delta_R^{(1)}U$ and $\delta_R^{(1)}\rho$ are of
the same order in $\Omega$.  The continuity equation,
(\ref{e:d1continuity}), then implies that $\deltaIRR \vec v = {\cal
  O}(\Omega\deltaIRR \rho)$.  Finally, the $\phi$-component of the
Euler equation gives, to lowest order in $\Omega$,
\be \delta_R^{(1)} U = 
\delta^{(1)}_\perp{\cal F}
+ {\cal O}(\Omega^2\deltaIRR\rho). 
\label{e:deltaU}\ee 
From its definition in Eq.~(\ref{e:Fgradient}) it follows that
$\delta^{(1)}_\perp{\cal F} ={\cal O}(\Omega\beta)$, which then implies that
$\deltaIRR\rho= {\cal O}(\beta\Omega)$ and $\delta_R^{(1)} \vec v = {\cal
  O}(\beta \Omega^2)$.

Dias and S\'a~\cite{Sa2005b} find, for an $\ell = 2$ perturbation, 
a solution $\deltaIRR \vec v, \deltaIRR U$ that is a sum of a) 
 our solution
with $ \deltaIRR U$ given by Eq.~(\ref{e:deltaU}) and b) 
a solution to the homogeneous equations with $\phi$-parity 
opposite to that of the Newtonian $r$-mode $\deltaINR Q$.  
As noted above, adding part b of their solution 
is equivalent to changing the initial phase of the perturbation.

\subsection{Second Order Axisymmetric Perturbations}  
\label{s:SecondOrderAxisymmetric}

In computing the quadratic terms that enter the second-order 
perturbation equations, it will be useful to have explicit expressions 
for the classical $r$-mode $\deltaINR v^a_1$ in cylindrical coordinates
$(\varpi,z,\phi)$,
\bsube
\bea
\!\!\!\!\!\!\!\!
\deltaINR v^\varpi_1&=&-\Omega_0\,z \left(\frac{\varpi}{R}\right)^{\ell-1} 
\cos(\ell\phi+\omega t),\\
\!\!\!\!\!\!\!\!
\deltaINR v^z_1&=&\Omega_0 \,R \left(\frac{\varpi}{R}\right)^{\ell}
\cos(\ell\phi+\omega t),\\
\!\!\!\!\!\!\!\!
\deltaINR v^\phi_1&=&
\Omega_0\,\frac{z}{R}\left(\frac{\varpi}{R}\right)^{\ell-2}
\!\!\sin(\ell\phi+\omega t).
\eea\label{e:deltav_cylindrical}\esube
From these one finds explicit expressions for the cylindrical components 
of the quadratic term $\bigl\langle \deltaINR v^b_1\nabla_b \deltaINR
v^a_1\bigr\rangle$, which appears as a source in the second-order Euler
equation, Eq.~(\ref{e:PerturbedEulerII}):
\bsube
\begin{align}
\bigl\langle\deltaINR \vec v_1\cdot\vec\nabla\deltaINR v^\varpi_1\bigr\rangle
 &=\frac{\Omega_0^2}{2R}\left[2(\ell-1)z^2-\varpi^2\right]
\left(\frac{\varpi}{R}\right)^{2\ell-3},
\\
\bigl\langle\deltaINR \vec v_1\cdot\vec\nabla\deltaINR v^z_1\bigr\rangle
  &=-\ell\Omega_0^2z \left(\frac{\varpi}{R}\right)^{2\ell-2},\\
\bigl\langle\deltaINR \vec v_1\cdot\vec\nabla\deltaINR v^\phi_1\bigr\rangle &=0.
\end{align}\esube

The axisymmetric parts of the nonradiative second-order perturbations
$\bigl\langle\deltaIINR v^a\bigr\rangle$ and $\bigl\langle\deltaIINR
U\bigr\rangle$ are determined by solving the perturbed Euler equation,
Eq.~(\ref{e:NRPerturbedEulerII}), and the perturbed mass conservation
equation, Eq.~(\ref{e:NRPerturbedMassConsII}).  
The contributions to each component of Euler's equation at lowest order in angular
velocity are given by,  
\bsube\bea 
0 = &&
\bigl\langle\deltaIINR E_\varpi\bigr\rangle =
-2\varpi\Omega_0\bigl\langle\deltaIINR v^\phi_1\bigr\rangle
+\partial_\varpi\bigl\langle\deltaIINR U_2\bigr\rangle\nonumber\\ 
&&\qquad\qquad
+\left[2(\ell-1)z^2-\varpi^2\right]\frac{\Omega_0^2}{2R}
\left(\frac{\varpi}{R}\right)^{2\ell -3},\qquad\quad
\label{e:PerturbedEulerVarpi}\\ 
0 = &&
\bigl\langle\deltaIINR
E_z\bigr\rangle = \partial_z\bigl\langle\deltaIINR U_2\bigr\rangle -\ell z
\Omega_0^2\left(\frac{\varpi}{R}\right)^{2\ell-2},
\label{e:PerturbedEulerZ}\\ 
0 = &&
\bigl\langle\deltaIINR
E_\phi\bigr\rangle = 2\varpi\Omega_0 \bigl\langle\deltaIINR v^\varpi_1
\bigr\rangle.
\label{e:PerturbedEulerPhi}
\eea \esube
The integrability conditions for these equations,
$\bigl\langle\deltaIINR E_{a}\bigr\rangle=0$, are given by
$\nabla_{[a}\bigl\langle\deltaIINR E_{b]}\bigr\rangle=0$.  In cylindrical coordinates, 
these integrability conditions, at lowest order in angular velocity are
\bsube
\bea
0&=&\nabla_{[z}\bigl\langle\deltaIINR E_{\varpi]}\bigr\rangle=
-\varpi\Omega_0\partial_z\bigl\langle\deltaIINR v^\phi_1\bigr\rangle\nonumber\\
&&
\qquad\qquad\qquad\quad
+(\ell^2-1)\frac{\Omega_0^2z}{R}\left(\frac{\varpi}{R}\right)^{2\ell-3},\qquad\\
0&=&\nabla_{[z}\bigl\langle\deltaIINR E_{\phi]}\bigr\rangle=
\Omega_0\partial_z\bigl\langle\deltaIINR v^\varpi_1\bigr\rangle,\\
0&=&\nabla_{[\varpi}\bigl\langle\deltaIINR E_{\phi]}\bigr\rangle=
\Omega_0\partial_\varpi\left(\varpi\bigl\langle\deltaIINR v^\varpi_1\bigr\rangle\right).
\eea\esube
These conditions, together with the requirement that the solution is 
nonsingular on the rotation axis, determine 
$\bigl\langle\deltaIINR v^\varpi_1\bigr\rangle$
and $\bigl\langle\deltaIINR v^\phi_1\bigr\rangle$, up to the time independent
differential rotation $\deltaIINR\Omega(\varpi)$
As before, we denote a particular choice by $\deltaII_{NP}v^\phi$: 
\bea
\bigl\langle\deltaIINR v^\varpi_1\bigr\rangle&=&0,
\label{e:deltaIINRVvarpi}\\
\bigl\langle\deltaII_{NP} v^\phi_1\bigr\rangle
  &=&(\ell^2-1)\frac{\Omega_0z^2}{2R^2}\left(\frac{\varpi}{R}\right)^{2\ell-4}.\quad
\label{e:delta2vphi}\eea
The remaining component, $\bigl\langle\deltaIINR v^z_1\bigr\rangle$, is
determined from the lowest order in angular velocity piece of the
perturbed mass conservation equation [cf. Eq.~(\ref{e:NRPerturbedMassConsII})],
\be
\nabla_a\left(\rho\bigl\langle\deltaIINR v^a_1\bigr\rangle\right)=0.
\ee
This equation, together with Eq.~(\ref{e:deltaIINRVvarpi}), shows that
the only nonsingular solution for $\bigl\langle\deltaIINR v^z_1\bigr\rangle$ is
\bea
\bigl\langle\deltaIINR v^z_1\bigr\rangle&=&0.
\eea

The scalar parts of the second order nonradiative $r$-mode,
$\bigl\langle\deltaIINR \rho\bigr\rangle$ and 
$\bigl\langle\deltaIINR \Phi\bigr\rangle$,
are determined by completing the solution to the perturbed Euler
equation $\bigl\langle\deltaIINR E_a\bigr\rangle=0$, and then solving the
perturbed gravitational potential equation.  The potential
$\bigl\langle\deltaIINR U\bigr\rangle$ is determined by integrating the
perturbed Euler Eqs.~(\ref{e:PerturbedEulerVarpi}) and
(\ref{e:PerturbedEulerZ}).  Using
Eqs.~(\ref{e:Omega_decomp}) and (\ref{e:delta2vphi}) we obtain the 
following expression for the axisymmetric part of the solution, to
lowest order in angular velocity,
\bea
\bigl\langle\deltaIINR U_2\bigr\rangle
&=&\frac{\Omega_0^2R^2}{4\ell}\left(\frac{\varpi}{R}\right)^{2\ell}
+\frac{\ell\,\Omega_0^2 z^2}{2}\left(\frac{\varpi}{R}\right)^{2\ell-2}\nonumber\\
&&+ 2\Omega_0\int^\varpi_0 \varpi'\deltaIINR\Omega(\varpi')d\varpi'
+ \deltaIINR C_2,\qquad
\label{e:deltaIIU2}
\eea
where $\deltaIINR C_2$ is a constant.     

The pressure as well as the density perturbations, $\deltaII p$ and
$\deltaII \rho$, are related to $\deltaII U$ as follows,
\bea
\deltaII U &=&\deltaII\Phi+\frac{1}{\rho}\deltaII p
   -\frac{1}{2\rho^2}\deltaI p\, \deltaI \rho \nonumber\\
&=&\deltaII\Phi+\frac{\gamma p }{\rho^2}\deltaII \rho
\nonumber\\
&&\quad+\frac{p}{2\rho^2}\left[\frac{\gamma(\gamma-2)}{\rho}
+\frac{d\gamma}{d\rho}\right](\deltaI \rho)^2,
\label{e:eosII}\eea
where $\gamma=d\log p/d \log \rho$ is the adiabatic index.  For the 
$r$-modes, the first-order perturbations $\deltaI p$ and $\deltaI \rho$ are ${\cal O}(\Omega^2)$.  So at lowest order in angular velocity, the 
relation between $\deltaII U$ and
$\deltaII\rho$ simplifies to
\bea
\deltaII U_2 &=&\deltaII\Phi_2+\frac{\gamma p }{\rho^2}\deltaII \rho_2.
\eea

The gravitational potential $\deltaII\Phi$ is determined by solving
the perturbed gravitational potential equation,
\bea
\nabla^2\deltaII\Phi &=&  4\pi\deltaII\rho.
\eea
For the $r$-modes, to lowest order in the angular velocity, this
equation my be rewritten as
\bea \nabla^2\deltaII\Phi_2 +\frac{4\pi\rho^2}{\gamma
  p_0}\deltaII\Phi_2 &=&\frac{4\pi\rho^2}{\gamma p_0} \deltaII U_2.
\eea
Using the expression derived in Eq.~(\ref{e:deltaIIU2}) for the
axisymmetric part of $\deltaIINR U_2$, we find the general
equation for $\bigl\langle\deltaIINR \Phi_2\bigr\rangle$:
\bea 
&&\nabla^2\bigl\langle\deltaIINR\Phi_2\bigr\rangle 
+\frac{4\pi\rho^2}{\gamma  p_0}\bigl\langle\deltaII\Phi_2\bigr\rangle 
\nonumber\\
&&\quad=\frac{4\pi\rho^2}{\gamma p_0}\left\{ 
\frac{\Omega_0^2R^2}{4\ell}\left(\frac{\varpi}{R}\right)^{2\ell} 
+\frac{\ell\,\Omega_0^2 z^2}{2}\left(\frac{\varpi}{R}\right)^{2\ell-2}
\right.\nonumber\\
&&\qquad\qquad\left.
+ 2\Omega_0\int^\varpi_0 \varpi'\deltaIINR\Omega(\varpi')d\varpi'+ 
\deltaIINR C_2\right\}.\qquad
\label{e:SecondOrderGravPotenialEq}
\eea

Finally, we use Eq.~(\ref{e:Integral4}) to obtain an explicit formula
for the second-order differential rotation, $\deltaIINR\Omega(\varpi)$, in terms of the second-order radiation-reaction
force and the second-order velocity perturbation $\deltaIINR v^a$.  Of
the three terms on the right side of that equation, we will see that
the second and third are higher order in $\Omega$ than the first, and
we will evaluate the first term to leading order in $\Omega$.
 
We first use Eq.~(\ref{e:fGR}) to find an explicit form for the second-order 
radiation-reaction force $\bigl\langle \deltaIIRR\! \vec f_{GR}\bigr\rangle$.
From Eqs.~(\ref{e:deltav_cylindrical}) and (\ref{e:deltaS}) for $\deltaINR
v^\theta$ and $\deltaINR S^{\ell\ell} $, we find
\be \bigl\langle \deltaIIRR\! \vec f_{GR}\bigr\rangle =
-\frac{(\ell+1)^2}4 \beta \Omega \left(\frac\varpi
R\right)^{2\ell-2}\vec\phi.  
\label{e:fGRSimplified}
\ee 
The second term $\delta_N^{(2)}v^\phi$ in Eq.~(\ref{e:betaF}) is given
by Eq.~(\ref{e:delta2vphi}). In the final term, $\delta_R^{(2)} V^\phi$, 
by its definition~(\ref{e:Vdef}), 
is proportional to a product of components of $\delta_N^{(1)} \vec v$ 
and $\delta_R^{(1)} \vec v$.  By our initial normalization, 
$\delta_N^{(1)} \vec v = {\cal O}(\Omega)$, 
and we found in Sect.~\ref{s:slowfirstorder} that  $\delta_R^{(1)} \vec v$ is 
${\cal O}(\Omega\delta_R^{(1)}\vec f_{GR}) = {\cal O}(\beta\Omega^2)$. 
 
From Eqs.~(\ref{e:betaF}), (\ref{e:fGRSimplified}), and
(\ref{e:delta2vphi}), we have
\bea
   \bigl\langle \deltaIIRR F^{\,\phi}\bigr\rangle  
	&=& -\Omega \left(\frac\varpi R\right)^{2\ell-4}
\left[ \frac{(\ell+1)^2}4 \left(\frac\varpi R\right)^{2} \right.
\nonumber\\
	&&\qquad
\qquad\quad\qquad\left.  + (\ell^2-1) \left(\frac zR\right)^2\right].\qquad
\label{e:Fphi}\eea
Equation~(\ref{e:Fphi}) implies $\bigl\langle \deltaIIRR
F^{\,\phi}\bigr\rangle ={\cal O}(\Omega)$.  The second term in
Eq.~(\ref{e:Integral4}) has integrand proportional to
$\mbox{$\bigl\langle \deltaIINR\rho\bigr\rangle$}$.
Because $\delta_N^{(2)}\rho = {\cal O}(\Omega^2)$, the integrand 
is ${\cal O}(\Omega^2)$, and the term itself is ${\cal O}(\Omega^3)$, 
two orders higher than $\bigl\langle \deltaIIRR F^{\,\phi}\bigr\rangle$.
Finally, the last term in (\ref{e:Integral4}) is proportional to
$\Omega\bigl\langle\deltaIIRR W^a\bigr\rangle$.  Eq.~(\ref{e:Wdef}) 
implies $\bigl\langle\deltaIIRR W^a\bigr\rangle = {\cal O}(\Omega^2)$, 
whence the last term is again ${\cal O}(\Omega^3)$.

With the dominant term in Eq.~(\ref{e:Integral4}) determined by
$\bigl\langle \deltaIIRR F^{\,\phi}\bigr\rangle$, we have
\be
\hspace{-4mm} \deltaIINR\Omega(\varpi)
 =\frac{\int_{-z_S}^{z_S} dz\, \rho\, \bigl\langle \deltaIIRR F^{\,\phi}\bigr\rangle}
	{2\int_{-z_S}^{z_S} dz\, \rho}. 
\label{e:d2ROmega}\ee
This integrand can be rewritten in a more explicit form using
Eqs.~(\ref{e:Fphi}) and (\ref{e:delta2vphi}):
\bea
\deltaIINR \Omega(\varpi)&=&
-\Omega\left(\frac\varpi R\right)^{2\ell-4}
\left[ \frac{(\ell+1)^2}8 \left(\frac\varpi R\right)^{2}\right.\nonumber\\
&&\qquad\qquad\qquad\quad \left.
+\frac{(\ell^2-1)}{2}\Upsilon(\varpi)\right],\qquad
\label{e:IIRROmega}
\eea
where $\Upsilon(\varpi)$ is the equation-of-state dependent, 
mass-weighted average of $(z/R)^2$,
\bea
\Upsilon(\varpi)&=&\frac{ \int_{-z_S}^{z_S} dz\, \rho z^2}
			{ R^2\int_{-z_S}^{z_S} dz\, \rho}.
\label{e:UpsilonDef}
\eea
The limits of integration, $\pm z_S(\varpi)$, in this expression  are the
$\varpi$ dependent values of $z$ at the surface of the equilibrium
star.  To lowest order in $\Omega$ these limits are the same as those
in a spherical nonrotating star: 
\be z_S(\varpi) =
\sqrt{R^2-\varpi^2}.  \ee 
The part of the second-order differential rotation that is not
explicitly caused by the radiation-reaction force, $\bigl\langle\deltaII_{NP} v^\phi_1\bigr\rangle$, is given in Eq.~(\ref{e:delta2vphi}):
\bea
\bigl\langle\deltaII_{NP} v^\phi\bigr\rangle
&=&(\ell^2-1)\frac{\Omega}{2}\left(\frac{ z}{R}\right)^2\left(\frac{\varpi}{R}\right)^{2\ell-4}.\quad
\label{e:IINROmega}
\eea
Together Eqs.~(\ref{e:IIRROmega}) and (\ref{e:IINROmega}) determine
(to lowest order in $\Omega$) the time-dependent differential rotation 
induced by gravitational-radiation reaction:
\bea
 \hspace{-4mm}  \delta^{(2)}\Omega 
&=& \left[\bigl\langle\deltaII_{NP} v^\phi\bigr\rangle+\deltaIINR \Omega(\varpi)
\right]e^{2\beta t}.\qquad
\label{e:d2Omega}\eea  

   The key result of this section is the derivation of an 
explicit expression (\ref{e:d2ROmega}) for 
$\deltaIINR\Omega(\varpi)$ in terms of the first-order $r$-mode. 
An expression of this kind exists because the rest of the 
second-order perturbation, the perturbed density, pressure, 
and potential, are higher-order in $\Omega$.  Like the 
velocity field of the first-order $r$-mode, the second-order 
differential rotation of the unstable $r$-mode can be found 
without simultaneously solving for the perturbed density 
and pressure.  

This separation of orders also leads to an iterative method 
for solving the second-order Newtonian perturbation equations at 
successive orders in $\Omega$ that mirrors the method we have just used to 
determine the axisymmetric parts of $\deltaIINR v^a$ at ${\cal O}(\Omega)$ and 
$\deltaIINR \rho$, $\deltaIINR p$, and $\deltaIINR \Phi$ 
at ${\cal O}(\Omega^2)$.  At each order, the ambiguity in the 
Newtonian differential rotation is resolved by using Eq.~(\ref{e:Integral4}).  
We assume that the first-order Newtonian perturbation 
equations have been solved to the desired order in $\Omega$.
We suppose one has found the perturbed Newtonian velocity 
$\deltaIINR v^a$ to ${\cal O}(\Omega^{2k-1})$ and the scalar quantities 
in $\deltaIINR Q$ to ${\cal O}(\Omega^{2k}$), and we list the steps 
to obtain the next-order correction: to find $\deltaIINR v^a_{2k+1}$
and the scalar quantities to ${\cal O}(\Omega^{2k+2})$.
    
\begin{enumerate}
\item Because $\deltaIINR v^a_{2k-1}$ is known,
and the integrability conditions  $\nabla_{[a}\deltaIINR E_{b]}=0$
have an additional power of $\Omega$ in each term, they are satisfied to 
at ${\cal O}(\Omega^{2k})$. One can then integrate the $\varpi$ or $z$ 
component of the perturbed Newtonian Euler equation (\ref{e:NRPerturbedEulerII}) 
to find $\deltaIINR U_{2k+2}$ up to a constant $\deltaIINR C_{2k+2}$. 
\item Equation~(\ref{e:eosII}) determines 
$\deltaIINR \rho_{2k+2}$ up to the ambiguity associated with 
$\deltaIINR C_{2k+2}$.
The Poisson equation, Eq.~(\ref{e:PerturbedPoissonII}), with the 
conditions that $\deltaIINR\Phi_{2k+2}$ vanish at infinity and have 
no monopole part (no change in mass), determines both $\deltaIINR\Phi_{2k+2}$ 
and the constant $\deltaIINR C_{2k+2}$.     
\item Equation~(\ref{e:eosII}) (or, alternatively, the Poisson equation) gives 
$\deltaIINR \rho_{2k+2}$, and the equation of state determines $\deltaIINR p_{2k+2}$. 
\item Finally, one uses the known first-order perturbation $\deltaINR v^a$ to solve 
two independent components of the curl of the Euler equation, 
$\deltaIINR E_{a}=0$ for $\deltaIINR v^\phi_{2k+1}$ and  
$\deltaIINR v^\varpi_{2k+1}$; $\bigl\langle\deltaIINR v^\phi_{2k+1}\bigr\rangle$ 
has an $f(\varpi)$ 
ambiguity that is resolved by Eq.~(\ref{e:Integral4}). 
The final component $\deltaIINR v^z_{2k+1}$ is found from the second-order 
mass-conservation equation. 
\end{enumerate}

\subsection{Secular drift of a fluid element}

The differential rotation we have found for the unstable $r$-mode
extends the work of S\'a and
collaborators~\cite{Sa2004}-\cite{Sa2005b} to obtain the differential
rotation of the unstable second-order $r$-mode.  The studies of
magnetic field wind-up by Rezzolla, {\it et al.}~\cite{Rezzolla00,Rezzolla01b,Rezzolla01c}, 
which predated this
work, explicitly omitted the form of the second order perturbation to
the velocity field that we have computed here.  These authors obtained
a secular drift $\phi(t)$ in the position of a fluid element by
integrating the $\ell=2$ form of the equations for the position
$\phi(t)$ and $\theta(t)$ of a particle whose perturbed velocity field
is found solely from the first-order perturbation $\deltaINR v^a$ of
Eq.~(\ref{e:ClassicalRModeVelocity}), from the equations \bsube \bea
\frac{d\theta}{dt} &=& \alpha\deltaINR v^\theta[\theta(t),\phi(t)],
\\ \frac{d\phi}{dt} &=& \alpha\deltaINR v^\phi[\theta(t),\phi(t)].
\label{e:drift}\eea\esube
The equations are nonlinear in $\theta(t), \phi(t)$, and the solution 
is written to ${\cal O}(\alpha^2)$.  The axisymmetric part of the solution 
is again the part that is not oscillatory in time; in our notation, it has the form
\be
\langle\theta(t)\rangle = 0, \quad 
\langle\phi(t)\rangle = \alpha^2 \frac34 \left[\left(\frac\varpi R\right)^2 - 2\left(\frac z R\right)^2\right] \Omega t.  
\ee
A secular drift obtained in this way has been used in subsequent papers by 
Cuofano, {\it et al.}~\cite{Cuofano2010,Cuofano_etal12}, and by 
Cao, {\it et al.}~\cite{CZW15}.  

When one includes the second-order differential rotation
$\deltaII\Omega$ of the unstable $\ell=2$ $r$-mode from
Eqs.~(\ref{e:d2Omega}), additional terms are
added to the secular drift $\phi(t)$ of a fluid element's position. 
The resulting expression is
given for $t\ll1/\beta$ by 
\be \langle\phi(t)\rangle = \alpha^2
\left\{\frac34 \left[\left(\frac\varpi R\right)^2 - 2\left(\frac z
  R\right)^2\right] \Omega+\deltaII\Omega|_{t=0}\right\} t. 
\ee 
Using the expression for $\deltaII\Omega$ in Eq.~(\ref{e:d2Omega}), with
Eqs.~(\ref{e:IIRROmega}) and (\ref{e:IINROmega}), we obtain the following 
explicit form for the second-order drift of an unstable $\ell=2$ $r$-mode:
\be
\langle\phi(t)\rangle = -\frac32 \alpha^2\,\Omega
\left[\frac14 \left(\frac\varpi R\right)^2+\Upsilon(\varpi)\right] t.
\label{e:phi_drift}\ee
This expression for the drift $\langle\phi(t)\rangle$ is independent
of $z$, and therefore describes a drift that is constant on
$\varpi\,=$ constant cylinders. The analogous expression for the drift
found previously by S\'a~\cite{Sa2004} has this same feature, and
Chugunov~\cite{Chugunov2015} observes that the drift in these modes
can therefore be completely eliminated in the pure Newtonian case by
appropriately choosing the arbitrary second-order angular velocity
perturbation.  

For long times (that is, for $\beta t$ arbitrary but
$\beta\ll\Omega$), the time dependence $t$ in Eq.~(\ref{e:phi_drift}) is replaced by\\
$\displaystyle (e^{2\beta t}-1)/2\beta$. This expression is not of
order $1/\beta$, but satisfies the bound 
\be 
	\frac{e^{2\beta t}-1}{2\beta} < t \frac{e^{2\beta t}+1}2, 
\ee 
for $t>0$.

\section{Polytropic Stellar Models}
\label{s:PolytropicStellarModels}

\begin{figure}
\includegraphics[width=3.in]{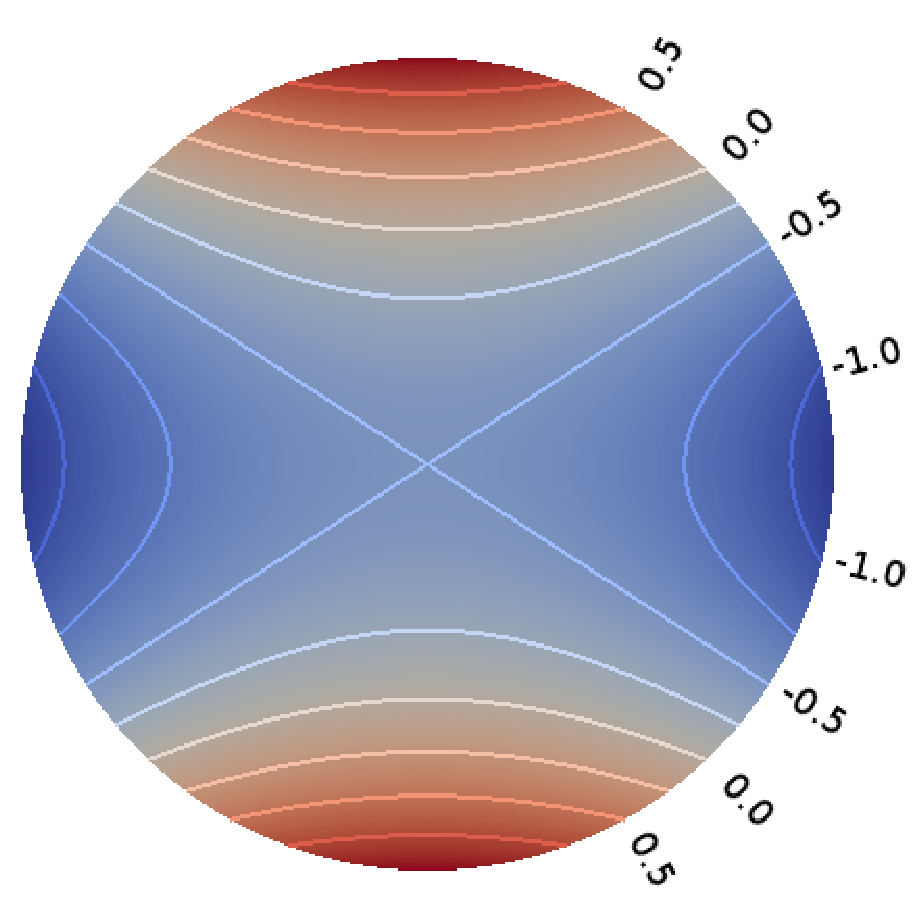}
\caption{\label{f:dOmegaContour} Differential rotation 
$\deltaII \Omega/\Omega$ from the $\ell=2$ $r$-mode instability
evaluated on a cross section through the rotation axis of a
slowly-rotating uniform-density star. The solution 
scales with time as $e^{2\beta t}$}
\end{figure}

In this section we evaluate Eq.~(\ref{e:d2Omega}), to determine the
changes in the rotation laws of uniformly rotating polytropes that are
induced by the gravitational-radiation driven instability in the
$r$-modes.  Polytropic stellar models (polytropes) are stars 
composed of matter whose equation of state has the form
\be
p = K\rho^{1+1/n},
\ee
where $K$ and $n$, the {\it polytropic index}, are constants.  
We start with the simplest case, $n=0$, the uniform-density
models. The only dependence of the differential rotation
$\delta^{(2)}\Omega$ on the equation of state is in
$\Upsilon(\varpi)$, the mass-weighted average of $(z/R)^2$ at 
fixed $\varpi$ defined in  
Eq.~(\ref{e:UpsilonDef}).  This average
can be evaluated analytically in the uniform-density case:
\bea
\Upsilon(\varpi)=\frac{R^2-\varpi^2}{3R^2} 
		= \frac{z_S^2(\varpi)}{3R^2}.
\eea
Combining this result with Eqs.~(\ref{e:IIRROmega}), (\ref{e:IINROmega})
and (\ref{e:d2Omega}), gives
\bea
 \hspace{-4mm} \delta^{(2)} \Omega 
&=& \Omega \left(\frac{\varpi}{R}\right)^{2\ell-4}
    \left[ \frac{(\ell+1)(\ell-7)}{24}\left(\frac{\varpi}{R}\right)^2\right.
\nonumber\\
	&&\hspace{15mm}\left.+ \frac{(\ell^2-1)}6\left(3\frac{z^2}{R^2}-1\right)
		\right]e^{2\beta t}.\qquad
\label{e:d20uniform}\eea  
In particular, for the $\ell=2$ $r$-mode, 
the radiation-reaction induced differential rotation has the form
\be \delta^{(2)}\Omega =
\Omega\left[\frac32\left(\frac{z}{R}\right)^2-\frac58
\left(\frac{\varpi}{R}\right)^2
-\frac12\right] e^{2\beta t},
\ee 
which is positive in a neighborhood of the poles and negative near the
equatorial plane.  Figure~\ref{f:dOmegaContour} illustrates the
gravitational-radiation driven differential rotation $\deltaII
\Omega/\Omega$ from the $\ell=2$ $r$-mode instability of a
slowly-rotating uniform-density star.  This figure shows contours of
constant $\deltaII \Omega/\Omega$, on a cross section of the star that
passes through the rotation axis.  For example, this figure ilustrates
that $\deltaII \Omega/\Omega\approx-9/8$ near the surface of the star
at the equator.  This indicates that the angular velocity of the star
is reduced by an amount $\approx-(9/8)\Omega\alpha^2 e^{2\beta t}$ in
this region, where $\alpha e^{\beta t}$ is the amplitude of the
$r$-mode, and $\Omega$ is the angular velocity of the unperturbed
star.  Similarly this figure illustrates that $\deltaII
\Omega/\Omega\approx 1$ near the poles.  The angular velocity of the
star is enhanced by the $r$-mode instability in these regions.

The equilibrium structures of $n=1$ polytropes can also be expressed
in terms of simple analytical functions, but the integrals that
determine $\Upsilon(\varpi)$ in Eq.~(\ref{e:UpsilonDef}) can
not. We therefore evaluate these quantities for all the $n\neq 0$
polytropes numerically.  

\begin{figure}
\includegraphics[width=3.3in]{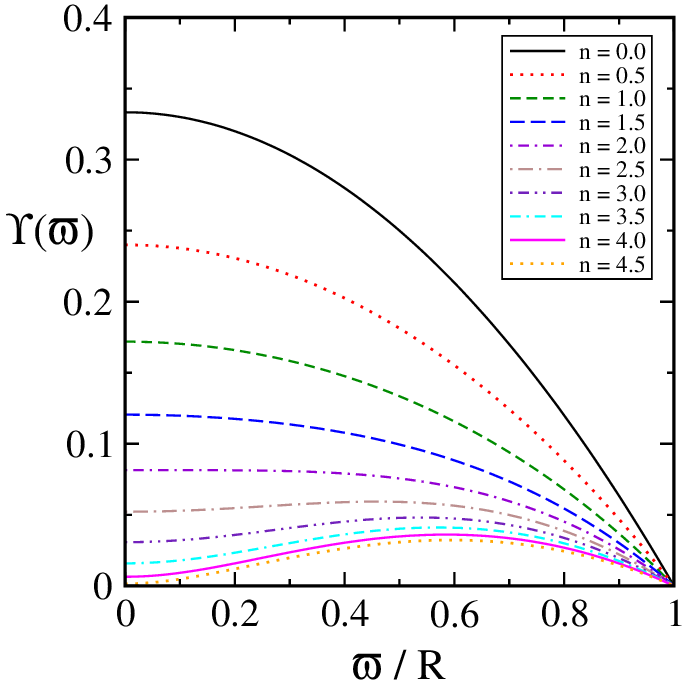}
\caption{\label{f:Upsilon} Dimensionless ratio of the integrals
  $\Upsilon(\varpi)$ defined in Eq.~(\ref{e:UpsilonDef}) that
  determines the gravitational-radiation induced differential rotation
  in polytropic stellar models having a range of polytropic indices
  $n$. }
\end{figure}

\begin{figure}
\includegraphics[width=3.3in]{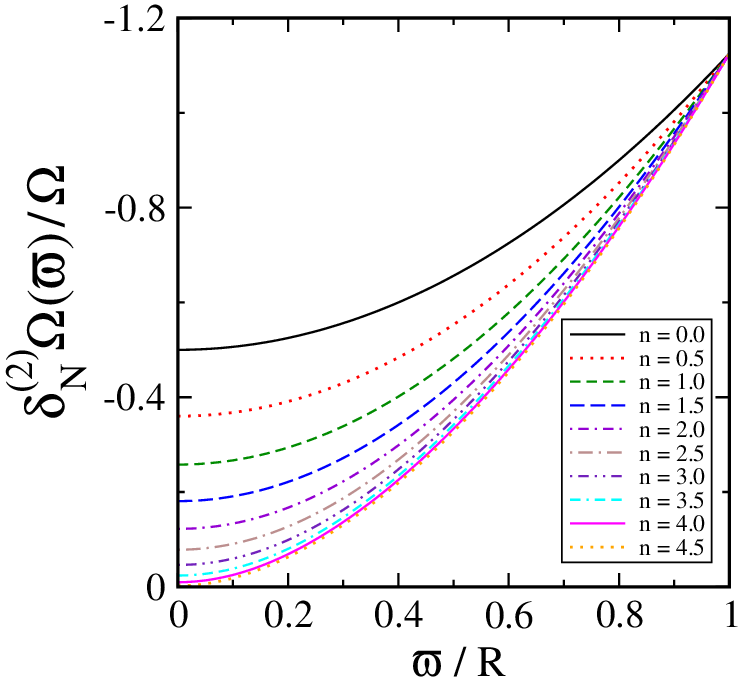}
\caption{\label{f:dOmega_n} Differential rotation induced by the
  gravitational-radiation instability in the $\ell=2$ $r$-modes for a
  range of polytropic indices $n$.}
\end{figure}

\begin{figure}
\includegraphics[width=3.3in]{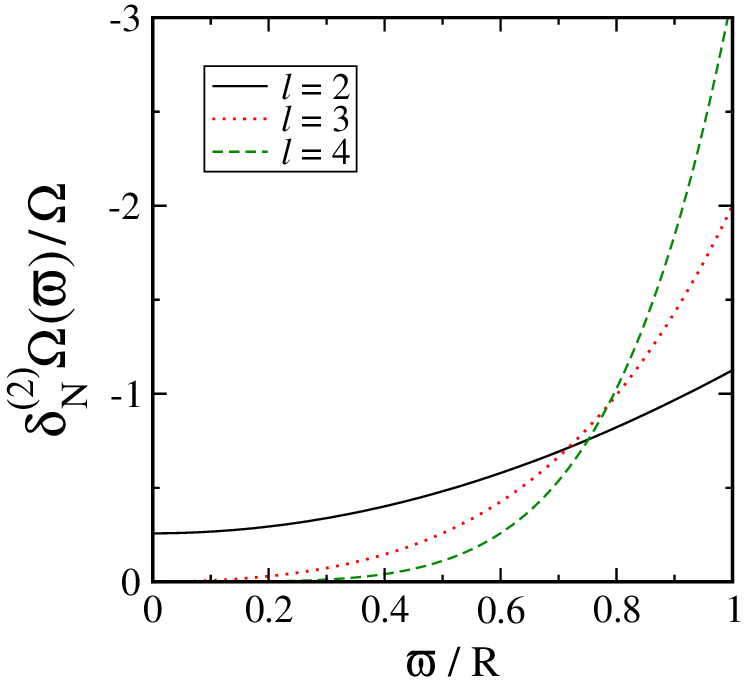}
\caption{\label{f:dOmega_L} Differential rotation induced by the
  gravitational-radiation instability in various $r$-modes 
of $n=1$ polytropes for a range of spherical harmonic
mode index $\ell$ values.}
\end{figure}

The structures of the non-rotating Newtonian polytropes are determined
by the Lane-Emden equations, which are generally written in the form,
\bea
\frac{d}{d\xi}\left(\xi^2\frac{d\theta}{d\xi}\right)=-\xi^2\theta^n,
\label{e:LaneEmden}
\eea
where $\theta$ is related to the density by $\rho=\rho_c\theta^n$,
with $\theta=1$ at the center of the star and $\theta=0$ at its
surface.  The variable $\xi$ is the scaled radial coordinate,
$r=a\xi$, with
\bea 
a^2=\frac{(n+1)K\rho_c^{(1-n)/n}}{4\pi G}.
\eea
We solve Eq.~(\ref{e:LaneEmden}) numerically to determine the
Lane-Emden functions $\theta(\xi)$, use them to evaluate the density
profiles of these stars, $\rho(r)=\rho_c\theta^n$, and finally perform
the integrals numerically in Eq.~(\ref{e:UpsilonDef}) that determine
the mass weighted average $\Upsilon(\varpi)$ of $(z/R)^2$ for 
spherical polytropes.  Figure~\ref{f:Upsilon} illustrates the results
for a range of polytropic indices.  Because they are more centrally 
condensed, stars with softer equations of state, i.e. polytropes with 
larger values of $n$, have smaller
$\Upsilon(\varpi)$. This is most pronounced 
near the rotation axis of the star where $\varpi=0$ and values of 
$z^2$ in the dense core dominate the average.  
Figure~\ref{f:dOmega_n} illustrates $\deltaIINR \Omega/\Omega$ from
Eq.~(\ref{e:IIRROmega}), the differential rotation induced by the
gravitational-radiation driven instability in the $\ell=2$ $r$-modes
for polytropes having a range of polytropic indices $n$.  This graph
shows that the equatorial surface value ($\varpi=R$) of $\deltaIINR
\Omega/\Omega$ is the same for all the polytropes.  This is not a
surprise, because $\Upsilon(\varpi)=0$ there for all equations of
state.  Stars composed of fluid having stiffer equations of state,
i.e. smaller values of $n$, have larger values of $|\deltaIINR
\Omega/\Omega|$ near the rotation axis where $\varpi=0$.
Figure~\ref{f:dOmega_L} illustrates the differential rotation induced
by the gravitational-radiation induced instability in the $r$-modes of
$n=1$ polytropes having a range of different spherical harmonic mode
index $\ell$ values.  The figure portrays a differential rotation 
$\deltaIINR \Omega/\Omega$ induced by gravitational radiation that, like the magnitude of the linear mode, is more
narrowly confined to the equatorial region near the surface of the
star as the $r$-mode harmonic index $\ell$ is increased.

\section{Discussion}

The radiation-reaction force uniquely determines the exponentially 
growing differential rotation of the unstable, nonlinear $r$-mode. 
We have found expressions for the rotation law and for the 
corresponding secular drift of a 
fluid element and have obtained their explicit forms for slowly rotating polytropes.
The formalism presented here describes an $r$-mode, driven by gravitational radiation reaction, at second order in its amplitude $\alpha$, and restricted to 
a perfect-fluid Newtonian model.  We now comment briefly on 
the meaning of the work within a broader physical context. 
 
First, a realistic evolution involves coupling to other modes, because
realistic initial data has small, nonzero initial amplitudes for all
modes and, at higher orders in $\alpha$, other modes are excited by
the $r$-mode itself.  As a result of the couplings, the $r$-mode
amplitude will saturate, and studies of its nonlinear evolution (see
\cite{Bondarescu07,Bondarescu09} and references therein) suggest a
saturation amplitude of order $10^{-4}$ or smaller.  By the time the
mode reaches saturation, the amplitude of daughter modes may be large
enough that their own second-order axisymmetric parts contribute
significantly to the differential rotation law.  

Second, when there is a background magnetic field, the growing axisymmetric 
magnetic field generated by the $r$-mode's secular drift can change the 
profile of the growing differential rotation \cite{Chugunov2015}.%
The second-order Euler equation 
(\ref{e:PerturbedEulerII}) is altered by the second-order Lorentz force per unit mass, 
given in an ideal magnetohydrodynamics approximation 
by 
 $\alpha^2\langle\deltaII f_{\rm magnetic}\rangle = \alpha^2\langle\deltaII [\frac1{4\pi\rho}(\nabla\times\vec B)\times \vec B]\rangle$.
This will be of order the radiation-reaction force after an amplitude-independent time%
\footnote{For a magnetic field that grows linearly in time, 
we have  
\[
   \alpha^2\langle\deltaII f_{\rm magnetic}\rangle \sim 
\alpha^2\frac1{4\pi\rho R} B_0^2\Omega t.
\]  The second-order radiation reaction force is given by  
$\alpha^2\deltaII f_{GR}\sim \alpha^2\beta\Omega R$, implying that the 
Lorentz force $\alpha^2\langle\deltaII f_{\rm magnetic}\rangle$ 
has comparable magnitude after a time given in Eq.~(\ref{e:tmagnetic}).
Here we follow Chugunov \cite{Chugunov2015}. 
Chugunov uses this argument to conclude that the magnetic field will not be significantly 
enhanced after it reaches $B \sim 10^8(\alpha/10^{-4})^2$ G, 
but his analysis is restricted to the case where the gravitational 
radiation-reaction force on the $r$-mode is negligible.
We have checked the conclusion of continued growth for Shapiro's model of 
a uniform-density cylinder with an initial magnetic field \cite{shapiro00}, 
by adding a forcing term of the form of the second-order axisymmetric 
radiation-reaction force~\cite{flr15}. We expect the amplification factor 
of the magnetic field to be limited by the value of the mode amplitude, 
$\alpha e^{\beta t}$, at nonlinear saturation, not by the value of the field, 
unless the initial magnetic field is of order $10^{12}$ G or larger. } 
\be
   t \sim \beta t_A^2 \sim 10^6 {\rm s}\ \frac{\rho}{10^{15}\rm g/cm^3} 
\frac{\beta}{10^{-6}{\rm s}^{-1}} \left(\frac{10^8\rm G}{B_0}\frac{R}{10^6\rm cm}\right)^2, 
\label{e:tmagnetic}\ee
where $t_A$ is the Alfv\'en time associated with the background field,
$t_A = \sqrt{4\pi\rho}/B_0$.  After this time and until the mode
reaches its nonlinear saturation amplitude, we expect that the
radiation-reaction force will continue to drive growing differential
rotation. The functional form of this differential rotation,
   however, will be determined by both $\deltaII f_{GR}$
  and $\langle\deltaII f_{\rm magnetic}\rangle$.

After nonlinear saturation, we expect the growth of differential
rotation and of the magnetic field to stop within a time on the order
of the Alfv\'en time. This is because (1) the radiation-reaction force
is now time independent, and (2), with a background magnetic field,
there should no longer be a zero-frequency subspace of modes
associated with adding differential rotation.  Reason (2) means that
the differential rotation and the magnetic field at the time of mode
saturation become initial data for a set of modes whose frequencies
are of order the Alv\'en frequency.  The second-order axisymmetric
part of the $r$-mode after saturation becomes effectively a system of
stable oscillators driven by a constant force.  Such systems have no
growing modes, and therefore no secularly growing magnetic field.

The explicit form of the secular drift we obtain
is new, but its magnitude is consistent with that used in earlier work
\cite{Rezzolla00,Rezzolla01b,Rezzolla01c,Cuofano2010,Cuofano_etal12,CZW15}
that examines the damping of the unstable $r$-mode by this energy transfer 
mechanism. This damping mechanism becomes important
  whenever the rate of energy transfer to the magnetic field (by
winding up magnetic field lines or, for a superconducting region in a
neutron-star interior, by stretching magnetic-flux tubes or other
mechanisms), is comparable to the growth rate of the unstable
  $r$-mode.  Assuming the energy transferred to the magnetic field is
not returned to the $r$-mode and that a large fraction of the core is
a type II superconductor, Rezzolla et al. \cite{Rezzolla00} estimate
that the instability will be magnetically damped for a magnetic field
of order $10^{12}$ G.  As noted above, we expect this magnetic
  damping mechanism to play a role only if the
magnetic field reaches this $10^{12}$ G range 
prior to nonlinear saturation of the $r$-mode.
We think it likely that 
a limit on magnetic field growth imposed by saturation means that this 
field strength can be reached only if the initial field is not far 
below $10^{12}$ G. In addition, for an initial field of order $B\geq 10^{12}$ G 
or larger, if all axisymmetric perturbations that 
wind up the magnetic field have frequency higher than or of order 
the Alfv\'en frequency, we conjecture (based on the toy model mentioned 
in Foonote 4) that the enhancement of the  
magnetic field will be too small to damp the $r$-mode. 
  
Finally, if the magnetic field is large enough to significantly modify
the structure of the first order r-modes, all of the calculations here
would need to be modified.  Previous studies, however
\cite{MR02,R02,lee05,GA07,LJP10,CS13,ARR12}, find that field strength
$B \gtrsim 10^{14}-10^{15}$ G is needed to significantly alter the
linear $r$-mode of a star with spin greater than 300 Hz.  When the
viscous damping time is comparable to the gravitational-wave growth
time, one would also need to include viscosity in the 2nd-order
equations that determine the differential rotation.

\acknowledgments 

We thank Andrey Chugunov for helpful comments on an earlier draft of
this manuscript, Luciano Rezzolla and Chugunov for discussions of
magnetic field evolution, and the referee for a careful reading,
useful suggestions, and insight into the likely role of nonlinear
saturation in the evolution of the r-mode's magnetic field.  JF thanks
Shin'ichirou Yoshida for corrections and contributions to an early set
of notes.  LL thanks the Leonard E. Parker Center for Gravitation,
Cosmology and Astrophysics, University of Wisconsin at Milwaukee for
their hospitality during several visits during which much of the
research presented here was performed.  LL was supported at Caltech in
part by a grant from the Sherman Fairchild Foundation and by grants
DMS-1065438 and PHY-1404569 from the National Science Foundation. JF
was supported in part by grant PHY-1001515 from the National Science
Foundation.

\clearpage

\appendix

\section{Notation}
\label{s:Notation}
\vspace{-6mm}

The symbols in Table~\ref{tab:fonts} are listed by order of appearance in the 
paper, {\it starting with Sec.~\ref{s:newtonian}}.  We omit a few symbols that are used only 
where they are defined.   
\vspace{-9mm}
\begin{table}[h]
\caption{\label{tab:fonts}Notation}
\begin{ruledtabular}
\begin{tabular}{ll}
\textbf{Symbol} & \textbf{Meaning} \\
$Q$  		& the set of variables $\{\rho,\vec v,p,\Phi\}$ \\
$\rho$ & mass density \\
$\vec v$ & fluid velocity \\
$p$ & fluid pressure \\
$\Phi$ & Newtonian gravitational potential\\
$h$ & fluid specific enthalpy\\
$U$ & effective potential \\
$E^a$ 		& $E^a=0$ is the Newtonian Euler equation\\
$\vec f_{GR}$	& radiation-reaction force \\
$I^{\ell m}, S^{\ell m}$ & mass and current multipoles\\
$N_\ell$	& a constant defined in Eq.~(\ref{e:NlDef}) \\
$\Omega$ & fluid angular velocity\\
$\vec\phi$	& rotational symmetry vector $x\hat y - y\hat x$\\
$\alpha$	& dimensionless amplitude of $r$-mode \\
$\delta^{(1)} Q$ & first-order perturbation of Q:
		     $\left.\partial_\alpha Q\right|_{\alpha=0}$\\
$\delta^{(2)} Q$ &second-order perturbation of Q: $\frac12\left.\partial_\alpha^2 Q\right|_{\alpha=0}$\\
$\deltaINR,\ \deltaIINR$ & first- and second-order Newtonian perturbations \\
			& (no radiation reaction)\\
$\deltaINR\hat Q$	& $\varpi,z$ dependence of perturbation: Eqs.~(\ref{e:deltaIrho})--(\ref{e:deltaIPhi})\\
$\deltaIRR\hat Q$ 	& a correction in first-order perturbation due to  \\
			& radiation reaction  \\
$\delta^{(1)} Q_\pm$ & subscript $\pm$ denotes even ($+$) or odd ($-$) $\phi$-parity \\
		& under the diffeomorphism $ \phi\rightarrow 2\pi-\phi$\\
$\omega_N$	& frequency of Newtonian $r$-mode \\
$\psi_N$ 	& $\psi_N \equiv \omega_Nt+m\phi$\\
$P^a{}_b$	& projection operator orthogonal to $\vec\phi$: Eq.~(\ref{e:projection})\\
$\beta$		& imaginary part of frequency of unstable $r$-mode\\
$\langle\delta Q\rangle$
		& axisymmetric part of $\delta Q$ \\ 
$\deltaII_{NP} Q$ &2nd-order Newtonian perturbation with a\\
		& particular choice of $\deltaIINR\Omega(\varpi)$\\
$\deltaIINR\Omega(\varpi)$
		& arbitrary function of $\varpi$ in second-order\\
		&  Newtonian differential rotation\\ 
$\deltaII\Omega$ & second-order differential rotation, $\langle\deltaII v^\phi\rangle$\\
$\deltaIIRR Q\, e^{2\beta t}$  
		& radiative part of second-order perturbation\\ 
$\deltaIIRR\vec V$,$\deltaIIRR\vec W$  	
		& defined in Eqs.~(\ref{e:Vdef}) and (\ref{e:Wdef}) \\
$\langle\deltaIIRR\vec F\rangle$  
		& effective driving force for $\langle\deltaIIRR\vec v\rangle$:
		  Eq.~(\ref{e:Fdef})\\
$M_0,\ R$	& mass and radius of spherical stellar model\\
$\Omega_0$	& $\sqrt{M_0/R^3}$\\
$\widetilde\Omega$ 	& dimensionless angular velocity, $\Omega/\Omega_0$\\
$Q_n$		& part of $Q$ that is $n$th order in $\widetilde\Omega$: Eq.~(\ref{e:Qn})\\
$\delta^{(1)}_\perp \vec f_{GR}$
		& part of $\delta^{(1)} \vec f_{GR}$ orthogonal to $\deltaINR\vec v$ \\
$\delta^{(1)}_\perp{\cal F}$ 
		& function for which 
		$\delta^{(1)}_\perp\vec f_{GR} = \vec\nabla \delta^{(1)}_\perp{\cal F}$ \\

\end{tabular}
\end{ruledtabular}
\end{table}

\section{Gravitational Wave Energy and Angular Momentum Fluxes}
\label{s:RadiationReaction}

The expression for the radiation reaction force $\vec f_{GR}$
  given in Eq.~(\ref{e:GRForceDef}) was derived by constructing a
  force that reproduces the standard expressions for the
time averaged gravitational wave energy and angular momentum fluxes:
\bea 
&&
\hspace{-6mm}
\left\langle\!\!\!\left\langle\left.\frac{dE}{dt}
\right\rangle\!\!\!\right\rangle \right|_{GR}
= \left\langle\!\!\!\left\langle \int \rho\, \vec v
\cdot \vec f_{GR}\,d^3x\right\rangle\!\!\!\right\rangle,\nonumber\\ 
&&\!\!\!\!\!\!\!\!=-\sum_{\ell\geq
  2}\sum_{|m|\leq\ell} \frac{1}{ 32\pi}
\left\langle\!\!\!\left\langle \left|\frac{d^{\,\ell+1}I^{\ell
    m}}{dt^{\,\ell+1}}\right|^2
+\left|\frac{d^{\,\ell+1}S^{\ell m}}{dt^{\,\ell+1}}\right|^2
\right\rangle\!\!\!\right\rangle,\nonumber\\
&&\qquad 
\label{e:EnergyFlux}
\\
&&
\hspace{-6mm}
\left\langle\!\!\!\left\langle\left.\frac{d\vec J}{dt}
\right\rangle\!\!\!\right\rangle \right|_{GR}
= \left\langle\!\!\!\left\langle \int \rho\, \vec r\times
\vec f_{GR}\,d^3x\right\rangle\!\!\!\right\rangle,\nonumber\\ 
&&\!\!\!\!\!\!\!\!=\sum_{\ell\geq
  2}\sum_{|m|\leq\ell} \frac{1}{ 32\pi}
\Re\left\langle\!\!\!\left\langle \frac{d^{\,\ell}I^{*\ell
    m}}{dt^{\,\ell}}\frac{d^{\,\ell+1}\vec I^{\,\,\ell
    m}_B}{dt^{\,\ell+1}}
\right.\right.\nonumber\\
&&\qquad\qquad\qquad\qquad\quad
+\left.\left. \frac{d^{\,\ell}S^{*\ell
    m}}{dt^{\,\ell}}\frac{d^{\,\ell+1}\vec S^{\,\,\ell
    m}_B}{dt^{\,\ell+1}}
\right\rangle\!\!\!\right\rangle.
\label{e:AngularMomentumFlux}
\eea
The expression given here for the angular momentum flux,
Eq.~(\ref{e:AngularMomentumFlux}), is somewhat more compact than the
standard post-Newtonian expression (cf. Thorne~\cite{Thorne1980}
Eq.~4.23).  We express this flux in terms of the magnetic type mass
and current multipole moments, $\vec I^{\,\,\ell m}_B$ and $\vec
S^{\,\,\ell m}_B$, which we define as
\bea
\vec I^{\,\,\ell m}_B &=& 
N_\ell\sqrt{\ell+1}\int\rho\, r^\ell \,\vec Y^{\,*\ell m}_B d^3x,\\
\vec S^{\,\,\ell m}_B &=& 
\frac{2N_\ell}{\sqrt{\ell+1}}\int\rho\, r^\ell \,
(\vec v \cdot \vec r \times \vec \nabla)\vec Y^{\,*\ell m}_B d^3x.
\eea
These magnetic type mass and current mutipole moments can be expressed 
in terms of the standard $I^{\ell m}$ and $S^{\ell m}$:
\bea
\vec I^{\,\,\ell m}_B&=& 
-\frac{i}{2}\sqrt{(\ell-m)(\ell+m+1)}
I^{\ell\,m+1}(\hat x + i \hat y)\nonumber\\
&&-\frac{i}{2}\sqrt{(\ell+m)(\ell-m+1)}
I^{\ell\,m-1}(\hat x - i \hat y)\nonumber\\
&&-im
I^{\ell\,m}\hat z,\\
\vec S^{\,\,\ell m}_B&=& 
-\frac{i}{2}\sqrt{\frac{(\ell-m)(\ell+m+1)}{\ell(\ell+1)}}
S^{\ell\,m+1}(\hat x + i \hat y)\nonumber\\
&&-\frac{i}{2}\sqrt{\frac{(\ell+m)(\ell-m+1)}{\ell(\ell+1)}}
S^{\ell\,m-1}(\hat x - i \hat y)\nonumber\\
&&-\frac{im}{\sqrt{\ell(\ell+1)}}
S^{\ell\,m}\hat z,
\eea
where $\hat x$, $\hat y$ and $\hat z$ are unit vectors.  Both of
these expressions are based on the following identity for vector
spherical harmonics:
\bea
\vec Y^{\,\,\ell m}_B&=& 
\frac{i}{2}\sqrt{\frac{(\ell-m)(\ell+m+1)}{\ell(\ell+1)}}
Y^{\ell\,m+1}(\hat x - i \hat y)\nonumber\\
&&+\frac{i}{2}\sqrt{\frac{(\ell+m)(\ell-m+1)}{\ell(\ell+1)}}
Y^{\ell\,m-1}(\hat x + i \hat y)\nonumber\\
&&+\frac{im}{\sqrt{\ell(\ell+1)}}
Y^{\ell\,m}\hat z,
\eea
Using this transformation, Eq.~(\ref{e:AngularMomentumFlux})
reproduces the standard post-Newtonian expression
(cf. Thorne~\cite{Thorne1980} Eq.~4.23). The calculation needed
  to verify that the expression for the radiation reaction force $\vec
  f_{GR}$ given in Eq.~(\ref{e:GRForceDef}) satisfies the
  time averaged gravitational wave energy and angular momentum flux
  expressions given in Eqs.~(\ref{e:EnergyFlux}) and
  (\ref{e:AngularMomentumFlux}) is straightforward, but lengthy.

\section{Integrating $\delta\rho$ }
\label{s:surface}

  For rotating equilibrium stellar models having polytropic
    equations of state with polytropic index $n$, the density $\rho \propto
  (\mbox{distance to the surface})^n$ near the star's surface.
We assume here that the surface of the perturbed star is
smooth as a function of $\alpha$ and $\vec x$.
  Although the surface itself is smooth, the behavior of $\rho$ near $\rho=0$
  implies that $\nabla_a\rho$ diverges for $n<1$ and
  $\nabla_a\nabla_b\rho$ diverges for $n<2$. It follows that
$\deltaI\rho$ and $\deltaII\rho$ diverge because they involve first and second
  derivatives, respectively, of the unpertubed density.  We show, however,  that continuity and differentiability of 
the star's surface as a 
  function of $\vec x$ and $\alpha$  imply finiteness of
  the integrals $\int_{-\infty}^{\infty} \deltaI\rho\, dz$ and
  $\int_{-\infty}^{\infty} \deltaII\rho\, dz$, when $\deltaII\rho$ is
  regarded as a distribution.

We first verify the claimed behavior of $\rho$ for the 
unperturbed polytrope and then use the form of the 
Lagrangian perturbation of the enthalpy to 
deduce the behavior of $\deltaI\rho$ and $\deltaII\rho$ 
near the surface.  Denote by $z_S^\pm(\alpha,t,\varpi,\phi)$ 
the values of $z$ at the top and bottom parts of the surface 
of the perturbed star. We again introduce the polytropic function $\theta$,
related to the specific enthalpy by $\theta = \rho_o/[(n+1)p_o]\ h$.  
Then $\rho = \rho_o\theta^n\Theta(z_S^+-z)\Theta(z-z_S^-)$, 
where $\Theta(z_S^+-z)=1$ for $z_S^+ >z$
and $\Theta(z_S^+-z)=0$ for $z_S^{+} <z$.  
For the unperturbed rotating polytrope, $\theta$ is finite with
finite derivatives at the surface of the star.%
\footnote{For the unperturbed star, Eq.~(\ref{e:EquilibriumEuler}) implies 
$\theta = \rho_o/[(n+1)p_o] ({\cal E} - \Phi+\frac12\varpi^2\Omega^2)$, 
where $\cal E$ is the constant injection energy per unit mass.  
Caffarelli and Friedman prove that $\rho$ is H\"{o}lder continuous, 
$\rho\in C^{0,\alpha} (\mathbb R^3)$ \cite{CF80}, and the Poisson equation 
then implies $\Phi\in C^{2,\alpha}(\mathbb R^3)$. Thus $\theta$ has 
one-sided first and second derivatives at the surface.
} The lack of smoothness in $\rho$ at the surface thus arises from 
the fact that $n$ is not an integer. We now show for the perturbed 
polytrope that $\rho$ is again proportional to 
$(\mbox{distance to the surface})^n$ to second order in $\alpha$.

The vanishing of $\theta$  at the surface of the perturbed star is equivalent to the vanishing 
of the Lagrangian perturbation of $\theta$ at the unperturbed 
surface: 
\be 
\Delta\theta = 0, 
\ee
where 
\be \Delta\theta := \theta(\alpha,t,\vec x +\vec\xi) -\theta(0,t,\vec x),
\label{e:Delta}\ee
with $\vec\xi(\alpha,t,\vec x)$ the exact Lagrangian displacement -- a
vector from the position $\vec x$ of each fluid element in the
unperturbed star to its position $\vec x + \vec \xi$ in the perturbed fluid.  Our assumption that the
  surface changes smoothly as a function of $\alpha$ and $\vec x$ is
  then the requirement that $\vec\xi$ and its derivatives are smooth
  at the surface of the star.
Writing 
\be \vec\xi = \alpha \vec\xi^{(1)} + \alpha^2\vec\xi^{(2)} 
+\mathcal{O}(\alpha^3)\ee
and taking derivatives of (\ref{e:Delta}) with respect to $\alpha$, we
have 
\bsube \bea \deltaI\theta &=& \Delta^{(1)}\theta -
\xi^{(1)a}\nabla_a \theta,\\ \deltaII\theta &=& \Delta^{(2)}\theta -
\xi^{(2)a}\nabla_a \theta - \xi^{(1)a}\nabla_a\delta^{(1)}\theta
\nonumber\\ &&\phantom{xxxxx}
-\frac12\xi^{(1)a}\xi^{(1)b}\nabla_a\nabla_b\theta.  \eea \esube 
Then
$\deltaI\theta$ and $\deltaII\theta$ are finite at the
unperturbed surface, and, to second order
in $\alpha$, we can write for $\theta$ the Taylor expansion \be
\theta(\alpha,z,\varpi) = \partial_z\theta|_{z_S^+} (z-z_S^+)+{\cal
  O}(z-z_S^+)^2, \ee for $z<z_S^+$.  The corresponding expansion for
$\rho = \rho_0\theta^n$ is thus \be \rho(\alpha,z,\varpi) = \rho_0 (
-\partial_z\theta|_{z_S^+})^n (z_S^+ -z)^n +{\cal O}(z_S^+
-z)^{n+1}.\qquad
\label{e:rhoZ}\ee

We can now show directly that the integrals $\int_{-\infty}^{\infty}
\deltaI\rho\, dz$ and $\int_{-\infty}^{\infty} \deltaII\rho\, dz$ are
finite for polytropic equations of state with any polytropic index
$n>0$ for which the equilibrium star has a finite surface. More
precisely, they are finite everywhere except the equator, where the
range of integration vanishes.

For a given value of $\varpi$, we choose $Z^\pm$ with $0<Z^+<z_S^+$
and $0>Z^->z_S^-$ for all
$\alpha < \epsilon$, for some finite $\epsilon>0$.  We write the integral
as a sum of three parts, 
\be 
\int_{-\infty}^\infty \delta\rho\, dz 
	= \int_{Z^-}^{Z^+}\delta\rho\, dz + \int_{Z^+}^\infty \delta\rho\, dz + \int_{-\infty}^{Z^-}\delta\rho\, dz.  \ee 
In the first integral on the
right side, $\deltaI \rho$ and $\deltaII\rho$ are finite, so we need
only consider  $\int_{Z^+}^\infty \deltaI\rho\, dz$,   
$\int_{Z^+}^\infty\deltaII \rho\, dz$, and the corresponding integrals near 
the bottom part of the surface. Because the finiteness argument is identical 
for the integrals near $z_S^-$ and $z_S^+$, we consider the integrals near 
$z_S^+$.   \\
 
 We have
\bea
 \partial_\alpha\rho 
&=& \partial_\alpha\left\{[\rho_0 ( -\partial_z\theta|_{z_S^+})^n (z_S^+ -z)^n  \right.\nonumber\\
&&\left.\phantom{xxx}+{\cal O}(z_S^+ -z)^{n+1}]\Theta(z_S^+-z)\right\}
\nonumber\\
&=& \rho_0 ( -\partial_z\theta|_{z_S^+})^n \partial_\alpha[(z_S^+ -z)^n\Theta(z_S^+-z)] 
\nonumber\\
&&+ {\cal O}(z_S^+ -z)^n
\nonumber\\
&=& -\rho_0 \partial_\alpha z_S^+ (-\partial_z\theta|_{z_S^+})^n \partial_z[(z_S^+ -z)^n\Theta(z_S^+-z)] 
\nonumber\\
&&+ {\cal O}(z_S^+ -z)^n,
\label{e:drho}\eea
implying  
\bea
  \deltaI\rho &=& -\rho_0 [\xi^{(1)z} (-\partial_z\theta)^n]_{z_S} \partial_z[(z_S-z)^n\Theta(z_S-z)] 
\nonumber\\
&&+ {\cal O}(z_S -z)^n,
\label{e:deltaI_rho}\eea
where we have used the relation 
$\partial_\alpha z_S^+|_{\alpha=0} = \xi^{(1)z}|_{z_S}$. 
From Eq.~(\ref{e:drho}), we have
\bea
\partial_\alpha^2\rho
&=& -\rho_0 \partial_\alpha z_S^+ (-\partial_z\theta|_{z_S^+})^n \partial_\alpha\partial_z[(z_S^+ -z)^n\Theta(z_S^+-z)] 
\nonumber\\
&&+ {\cal O}(z_S^+ -z)^{n-1}
\nonumber\\
&=& \rho_0 (\partial_\alpha z_S^+)^2 (-\partial_z\theta|_{z_S^+})^n \partial_z^2[(z_S^+ -z)^n\Theta(z_S^+-z)] 
\nonumber\\
&&+ {\cal O}(z_S^+ -z)^{n-1},
\eea
implying
\begin{align}
  \deltaII\rho &= \frac12\rho_0 [(\xi^{(1)z})^2 (-\partial_z\theta)^n]_{z_S} \partial_z^2[(z_S-z)^n\Theta(z_S^+-z)] 
\nonumber\\
&\phantom{xx}+ {\cal O}(z_S-z)^{n-1}.
\label{e:deltaIIrho}\end{align}

Finiteness of $\int \deltaI\rho\,dz $ is immediate from the integrability
of $(z_S-z)^{n-1}$ for $n>0$.  For $\deltaII\rho$, we had to retain 
the factor $\Theta(z_S-z)$, and we kept it for $\deltaI\rho$ as well 
to display pairs of analogous equations.  
From Eqs.~(\ref{e:deltaI_rho}) and (\ref{e:deltaIIrho}), the leading term 
in each of $\deltaI\rho$ and $\deltaII\rho$ is a $z$-derivative, 
and we immediately obtain the integrals 
\bea
  \int_{Z^+}^\infty \deltaI\rho \, dz
&=& \rho(Z)\xi^{(1)z}|_{z_S}+ {\cal O}(z_S -Z^+)^{n+1},\qquad
\eea
\be
  \int_{Z^+}^\infty \deltaII\rho \, dz
  = \frac{n}2\frac{\rho(Z)}{z_S -Z^+}(\xi^{(1)z}|_{z_S})^2
     + {\cal O}(z_S -Z^+)^n.\qquad
\ee  
The integrals $\int_{-\infty}^\infty\deltaI \rho\,dz$ and 
$\int_{-\infty}^\infty\deltaII \rho\,dz$ are therefore finite as 
claimed.

\section{Ordering in $\Omega$ of $\deltaIRR Q$}
\label{s:ordering_appendix}
To make the heuristic argument of Sect.~\ref{s:slowfirstorder} more precise, we use the 
two-potential formalism of Ipser and Lindblom \cite{IL90} 
to write an explicit form 
for $\deltaIRR v^a$ in terms of $\deltaIRR U$ and $\deltaIRR {\cal F}$.  
Because that formalism uses the complex version of a perturbation, 
we write $\deltaIRR Q = \Im (\tdeltaIRR Q)$.
The perturbed Euler equation, Eq.~(\ref{e:d1Euler}), 
with radiation-reaction force then 
has the form  
\bea
  Q^{-1}_{ab} \tdeltaIRR v^b
    &\equiv& \left[(\omega_N+\ell\Omega) g_{ab} +2i\Omega\nabla_a \phi_b \right] \tdeltaIRR v^b
\nonumber\\
      &=& i\nabla_a\left(\tdeltaIRR U - \tdeltaIRR{\cal F}\right).  
\eea
Using the slow-rotation form (\ref{e:RModeFrequency}) of $\omega_N$ and 
Eq.~(13) of Ref.~\cite{IL90}, 
we write the solution to this equation for $\tdeltaIRR v^a$ as 
\be
   \tdeltaIRR v^a = iQ^{ab} \nabla_b\left(\tdeltaIRR U 
			- \tilde\delta^{(1)}_\perp{\cal F}\right),
\label{e:tildedv}\ee 
where the inverse of $Q^{-1}_{ab}$ is the tensor 
$Q^{ab} = \Omega^{-1} \widetilde Q^{ab}$, 
with 
\be
  \widetilde Q^{ab} = -\frac{\ell+1}{2\ell(\ell+2)}\left[g^{ab}
	  -(\ell+1)^2\nabla^a z\nabla^b z- i(\ell+1)\nabla^a\phi^b\right].
\ee

With $\tdeltaIRR v^a$ replaced by the expression on the right side of 
Eq.~(\ref{e:tildedv}), the mass conservation equation, 
Eq.~(\ref{e:d1continuity}) 
becomes an elliptic equation for $\tdeltaIRR U- \delta^{(1)}_\perp{\cal F}$, namely
\bea
 \hspace{-5mm} \nabla_a&&
\left[\rho \widetilde Q^{ab}\nabla_b\left(\tdeltaIRR U
	- \delta^{(1)}_\perp{\cal F}\right)\right] 
\nonumber\\
  && + \frac2{\ell+1} \Omega^2\rho \frac{d\rho}{dp} (\tdeltaIRR U - \tdeltaIRR\Phi)
    =  i\Omega\,\beta\,\deltaINR\rho.  
\label{e:U}\eea 
The potentials $\tdeltaIRR U$ and $\tdeltaIRR \Phi$ are determined by this equation, 
together with the Poisson equation,
\be
  \nabla^2\tdeltaIRR \Phi = 4\pi\rho\frac{d\rho}{dp}(\tdeltaIRR U - \tdeltaIRR\Phi),
\label{e:poisson}\ee
and the two boundary conditions, \be \lim_{r\rightarrow\infty}
\tdeltaIRR\Phi = 0 \ee and \be \DeltaI h = \left.\left(\tdeltaIRR U|_S
-\tdeltaIRR\Phi + \tilde\xi^{(1)\,b}_R\nabla_b h\right)\right|_S = 0;
\ee here $S$ is the surface of the unperturbed star and the Lagrangian
displacement $\tilde\xi^{(1)\,a}_R$ is defined by \be
\tilde\xi^{(1)\,a}_R = \frac1{i(\omega_N+\ell\,\Omega)} \tdeltaIRR
v^a.  \ee Using the value of
$\omega_N$ from Eq.~(\ref{e:RModeFrequency}), and
Eq.~(\ref{e:tildedv}), we can write the second boundary condition as
\be \widetilde Q^{ab}\nabla_a h \nabla_b (\tdeltaIRR U-\tilde
\delta^{(1)}_\perp{\cal F}) + \frac2{\ell+1} \Omega^2 (\tdeltaIRR U -
\tdeltaIRR\Phi) = 0.  \ee

To find the orders in $\Omega$ of $\deltaIRR \vec v$, $\deltaIRR U$ and 
$\deltaIRR\Phi$, we begin with the relations $\deltaINR\rho={\cal O}(\Omega^2)$
and, from Eq.~(\ref{e:Fgradient}), 
$\tilde\delta^{(1)}_\perp{\cal F} = {\cal O}(\Omega\beta)$.
From the Poisson equation (\ref{e:poisson}), $\deltaIRR U$ and $\deltaIRR\Phi$ are 
the same order in $\Omega$.    
From Eq.~(\ref{e:U}), we then have 
$\deltaIRR U - \delta^{(1)}_\perp{\cal F} 
  = {\cal O}(\Omega^2 \delta^{(1)}_\perp{\cal F})+{\cal O}(\Omega^3\beta)
	= {\cal O}(\Omega^3\beta)$.   
Then  
\be
  \deltaIRR\Phi = {\cal O}(\deltaIRR U) 
	=  {\cal O}(\delta^{(1)}_\perp{\cal F}) = {\cal O}(\Omega\beta).
\ee 
Finally, Eq.~(\ref{e:tildedv}) implies
\be
   \deltaIRR v^a = {\cal O}(\Omega^{-1}\Omega^3\beta) 
	= {\cal O}(\Omega^2\beta).
\ee


\bibstyle{prd} 
\bibliography{./References} 
\end{document}